\documentclass{article}

\usepackage[preprint]{neurips_2026}

\usepackage[utf8]{inputenc} 
\usepackage[T1]{fontenc}    
\usepackage{hyperref}       
\usepackage{url}            
\usepackage{booktabs}       
\usepackage{amsfonts}       
\usepackage{nicefrac}       
\usepackage{microtype}      
\usepackage[table,xcdraw]{xcolor}         
\usepackage{algorithm}
\usepackage[noend]{algpseudocode}
\usepackage{tikz}
\usepackage{pgfplots}
\usepackage{subcaption}
\usepackage{makecell}
\usepackage{listings}
\usepackage{graphicx}
\usepackage{multirow}
\pgfplotsset{compat=1.18}
\usepackage{wrapfig}
\usepackage{amsmath}
\usepackage{cleveref}
\usepackage{mathtools} 
\definecolor{stageblue}{HTML}{024995}
\definecolor{stagegreen}{HTML}{39732B}
\definecolor{stageorange}{HTML}{E06906}
\definecolor{stagepurple}{HTML}{5A3287}
\definecolor{stageyellow}{HTML}{E39824}
\usepackage{tabularx}

\usepackage{booktabs}
\newcommand{\MD}[0]{MathDial}
\newcommand{\GD}[0]{GeoDial}
\newcommand{\D}[0]{\textit{Decision}}
\newcommand{\Pd}[0]{\textit{Prediction}}
\newcommand{\titleimaged}{\protect\includegraphics[height=1em]{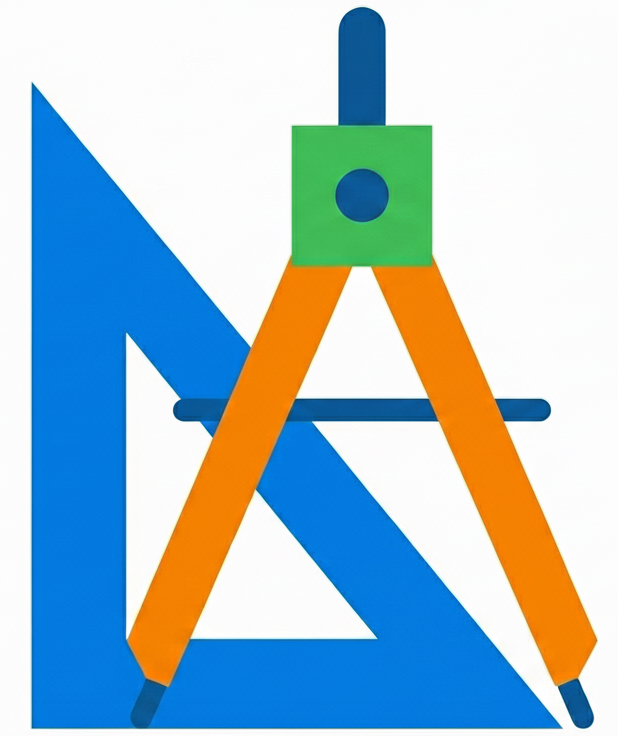}}
\title{\titleimaged\GD: A Multimodal Dialog Tutoring Dataset for Geometry Problem-Solving with Visual Tutor Turns}
\definecolor{actcolor}{RGB}{206, 255, 207}
\definecolor{hlcolor}{RGB}{225, 213, 232}
\definecolor{fbcolor}{RGB}{217, 232, 251}
\definecolor{genericcolor}{RGB}{81, 153, 199}
\definecolor{seefigcolor}{RGB}{253, 131, 46}
\definecolor{focuscolor}{RGB}{241, 68, 51}
\definecolor{probingcolor}{RGB}{73, 173, 93}
\definecolor{tellingcolor}{RGB}{138, 135, 180}
\definecolor{lightpink}{HTML}{E4DFF4}
\author{%
  Sankalan Pal Chowdhury\thanks{Equal contribution}\\
  ETH Zurich\\
  \texttt{sankalan.palchowdhury@inf.ethz.ch}
  \And
  Junling Wang\footnotemark[1] \\
  ETH Zurich, ETH AI Center\\
  \texttt{junling.wang@ai.ethz.ch}
  \AND
  Donya Rooein\\
  Bocconi University\\
  \texttt{donya.rooein@unibocconi.it}
  \And
  April Yi Wang\\
  ETH Zurich\\
  \texttt{april.wang@inf.ethz.ch}  
  \And
  Mrinmaya Sachan\\
  ETH Zurich\\
  \texttt{mrinmaya.sachan@inf.ethz.ch}
}

\begin{document}

\maketitle

\begin{abstract}
Several educational domains rely heavily on diagrams and visual cues, yet most existing tutoring datasets are limited to text-only interactions. This limits the development of AI tutors that can 
teach in visually grounded ways used by human instructors.
Thus, we introduce \GD, a multimodal tutoring dataset of over 1.3K teacher-student dialogs in the domain of geometry collected from experienced math teachers, where instructional turns are explicitly grounded in diagram highlights. 
We propose a scalable annotation protocol 
that integrates dialog acts, visual highlighting, and feedback, enabling fine-grained supervision of both language and visual tutoring behavior. To illustrate the challenges posed by this setting, we fine-tune several vision–language models on \GD\ and evaluate their ability to generate tutoring utterances and diagram highlights. While supervised fine-tuning substantially improves the quality of generated dialog, it struggles to produce accurate diagram highlights, revealing a key limitation of current methods 
and highlighting the need for approaches that more effectively integrate visual reasoning with pedagogical interaction.

\end{abstract}

\section{Introduction}


Students are often said to remember 10\% of what they read, 20\% of what they hear, and 30\% of what they see, a widely cited interpretation of Dale's ``Cone of Experience'' \citep{dale1969audiovisual}.
 Although specific percentages can be debated \citep{schneider2019much}, a large body of research emphasizes the critical role 
that visual information plays 
in learning across subjects \citep{bloom19842, topping2000tutoring} such as mathematics \citep{SCHOENHERR2024100639}. 
Indeed, it is hard to picture a ``classroom'' without a blackboard, or one of its high-tech replacements. For at least 2 centuries, the blackboard has acted as a visual mirror for mental processes \citep{phillips2015officer}.
\textbf{Yet, in the era of rapid adoption of Large Language Models (LLMs) in education, their pedagogy 
largely remains limited to the textual modality.}

Indeed, computer-based tutors that existed before LLMs, 
such as AutoTutor \citep{graesser2004autotutor} and Cognitive Tutor \citep{ritter2007cognitive}, also used both text and diagrams. However, they were hindered by scaling issues in authoring their scripts \citep{100hrs}. The ability of LLMs to distill knowledge in fluent natural language is seen as a way around these scaling issues, leading to the creation of several datasets \citep{macina2023mathdial, stasaski-etal-2020-cima, demszky-hill-2023-ncte, 10.1145/3698205.3733961, wang-etal-2024-bridging} recording different types of teacher-student interactions, which could then be used to train LLMs to be better teachers \citep{learnlmteam2025learnlmimprovinggeminilearning}. However, the initial LLMs had limited visual capabilities. 

With the rising capabilities of Vision-Language Models (VLMs), which can work with images in addition to text, there is an opportunity to bring back the visual modality in tutoring. 
A first step toward this goal is to develop multimodal tutoring datasets that capture how teachers coordinate language with visual guidance.
In this work, we propose \GD\footnote{The full dataset is made available via \href{https://kaggle.com/datasets/75a1dd70c8fc33d5bd862c7cd056cb38275521b018dec4aa5fde538fc25d5817}{Kaggle}}, a collection of over 1.3K richly annotated geometry tutoring dialogs that capture teacher responses to controlled, diagram-grounded student errors, including teachers' diagram annotations, visual highlights, and turn-level textual feedback.
This provides, to our knowledge, the first dataset for training and evaluating VLMs that support tutoring through coordinated language and visual guidance.


\begin{wrapfigure}{r}{0.5\textwidth}
    \vspace{-30pt}
    \includegraphics[width=0.49\textwidth]{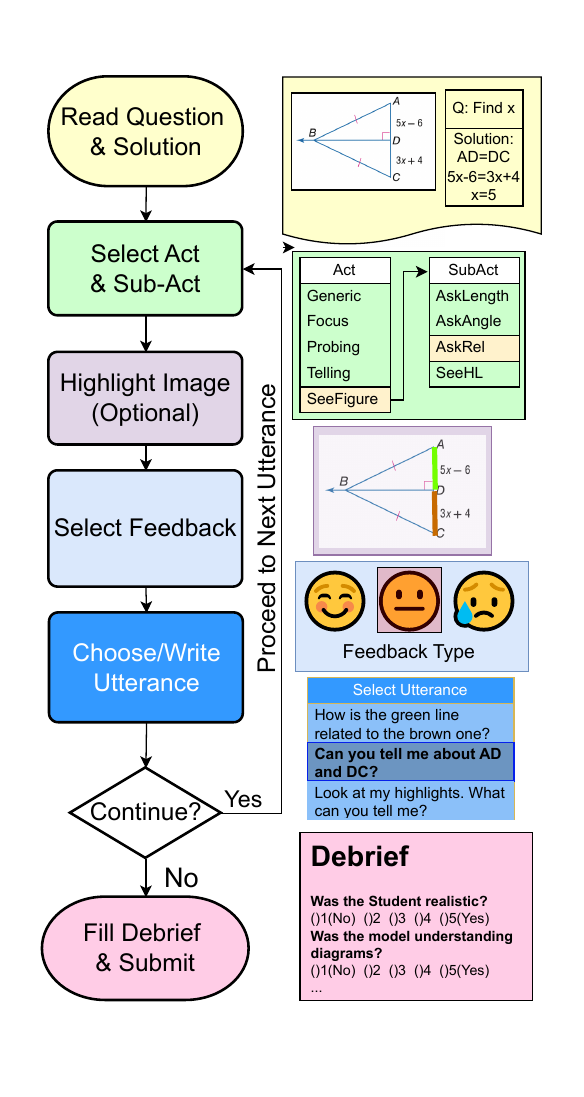}
    \vspace{-20pt}
    \caption{Flowchart showing our setup to collect dialogs. Surrounding infoboxes give examples of the corresponding step in the flowchart with the same color.}
    \vspace{-60pt}
    \label{fig:setup}
\end{wrapfigure}

Our experiments show that standard training on \GD\ improves pedagogical tutor-turn generation, but does not produce teacher-like diagram highlights: models learn to abstain more often, yet still struggle to select the correct visual elements. Further analysis suggests that this challenge stems from both highlight sparsity and the tight coupling between visual actions and teacher utterances. Focused highlight experiments show that separating highlight prediction from utterance generation and reweighting positive examples improves performance, suggesting that \GD\ exposes a challenging task for visually grounded pedagogical modeling. We open-source our dataset and models, with the aim of encouraging future work to overcome these limitations through better training and modeling paradigms.


\vspace{-2mm}
\section{Data Annotation Protocol}
\subsection{Annotation Process Overview}
\label{sec:anover}
Our goal is to gather tutoring conversations where a teacher 
is interacting one-on-one with a student. Good quality tutoring is a balancing act: offering a student too little help might lead to them getting bored and disengage, while offering too much support can cause them to become over dependent \citep{Reiser01072004,butler2013explanation}. The goal of the tutor is thus to guide students to the correct solution of a problem, without revealing too much of the solution themselves. Here, we focus on the corrective behaviour more than supportive behaviour.

We assume that the student has already made an attempt to solve a problem, and has gotten it wrong, before approaching the tutor. The conversation starts with the teacher seeing the question, the associated diagram, the correct solution and the student's solution. The student also has access to all of these, except the correct solution. 
The student, in our setup, is played by a VLM. We use VLM students as a controlled elicitation mechanism to collect teacher responses to plausible geometry errors under comparable conditions. While this weakens ecological validity, it avoids privacy and consent issues, speeds up data collection, and limits filler or generic utterances not specific to tutoring. Further, as we discuss in \Cref{sec:fid}, teachers judged the simulated students to exhibit strong fidelity to real student behavior.

\Cref{fig:setup} shows an overview of our setup. The conversation proceeds with the teacher and the student taking turns, until the teacher decides to end the conversation. Each utterance required the teachers to select an act and a subact, make optional highlights to the diagram and select a feedback type. The act and subact specify the teacher's strategy while the feedback is usually a single sentence prefix to the utterance, either acknowledging what they did (eg `Good job,' `That's right' etc.) or encouraging them to reconsider (eg. `Not quite.'). We discuss these in \Cref{sec:da,sec:fs} respectively. The highlights offer a visual channel of communication, which is the core contribution of \GD. We shall discuss them in \Cref{sec:hl}. All these act as guides for the actual utterance. 
Hereafter, three VLMs (Gemini-Flash-2.5 \citep{googleGeminiFlash}, GPT4o \citep{gpt}, and Claude-Sonnet-4 \citep{anthropic2025claude4systemcard}) suggest candidate utterances from the teacher-provided guides, leaving the annotator in control of the tutoring strategy. This process is repeated until the annotator wishes to end the conversation. A screenshot of the full annotation interface is provided in \Cref{fig:interface}. After each conversation, annotators complete a debrief form (details in \Cref{sec:debrief}). 

Using VLMs to suggest teacher utterances also poses a challenge to ecological validity. However, suggesting potential utterances to the teacher works to reduce their workload and resulting fatigue, and previous work has shown that fatigue can affect tutoring quality to the extent that human tutors are perceived as worse than AI tutors \citep{pal-chowdhury-etal-2025-educators}. We note that these suggestions are still based on the guides provided by the teacher so they remain in control of the overarching strategy. Also, they can always choose to ignore suggestions and write their own utterance.

\begin{wrapfigure}{r}{0.6\textwidth}
    \centering
    \vspace{-50pt}
    \includegraphics[width=0.59\textwidth, trim={0 2cm 2.5cm 1.5cm}, clip]{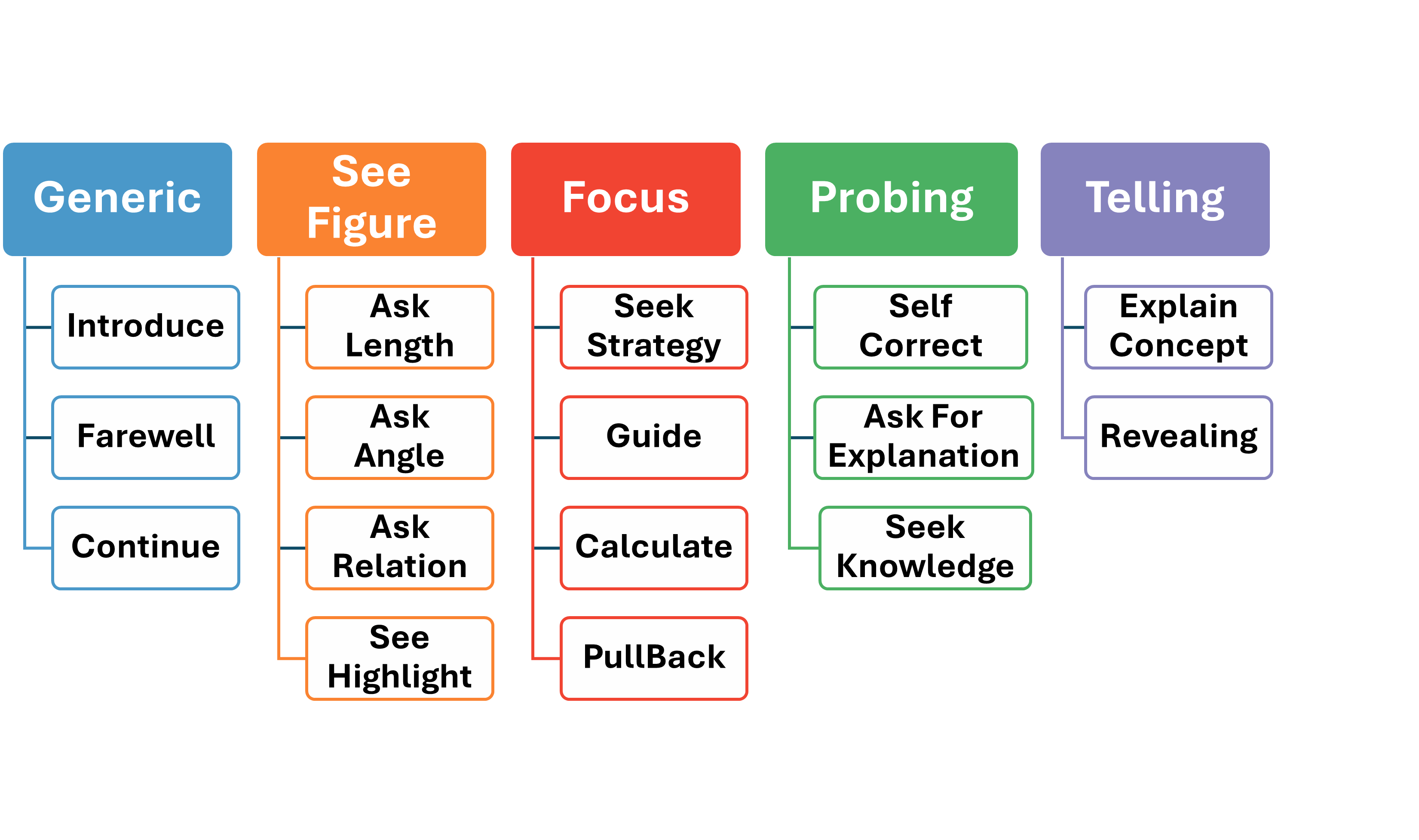}
    \caption{Tutor strategies (acts and subacts) in \GD.}
    \label{fig:subacts}
    \vspace{-10pt}
\end{wrapfigure}
\subsection{\colorbox{actcolor}{Dialog Acts}}\label{sec:da}
Dialog acts are the first step to specify the intent of the annotator, and thus need to be specific in order for the VLMs to generate the best possible suggestions. To aid the annotators in easily selecting their intent, we define a two-level hierarchy of $5$ acts and $16
$ subacts. Four of our acts are taken from \MD\ \citep{macina2023mathdial}, namely \textit{\color{focuscolor}Focus}, which are utterances making direct progress towards the solution, \textit{\color{probingcolor}Probing}, which tests the student's understanding of underlying concepts, \textit{\color{tellingcolor}Telling}, which reveals the solution or parts of it and \textit{\color{genericcolor}Generic}, which covers non-pedagogical conversational elements. To this we add a fifth act, \textit{\color{seefigcolor}SeeFigure}, which deals with the visual component of our task. These acts are then split into 16 subacts as shown in \Cref{fig:subacts}. The subacts for the \MD\ acts are mostly sourced from the intents they define in Table 2 of their paper. Each subact is associated with a prompt given to each of the 3 proposer VLMs. The full list of subacts, their descriptions, and examples is provided in \Cref{tab:subacts}.

\subsection{\colorbox{hlcolor}{Highlighting the Diagram}}\label{sec:hl}
Highlighting the diagram is an important part of the task, and is mainly used to get the student to focus on certain parts of the diagram. Once the annotator has selected the act and subact, they can proceed to highlight the diagram. The diagram is present in the top left corner of their screen (see  \Cref{fig:interface}), and the annotators are provided with a pen tool with which they can draw on the diagram. The color of this pen is fixed to green, as it is the rarest on the diagrams of the basic English colors, thereby unlikely to be confused with parts of the base diagram by the student model. An Eraser tool is also provided to fix mistakes. Additionally, we allow annotators to clear all their highlights and start fresh, or reuse the highlights from the previous step if they wish.

While annotators are allowed to highlight for all acts and subacts, some subacts, like \textit{Farewell} and \textit{PullBack}, are unlikely to need highlights. On the other hand, the \textit{\color{seefigcolor}SeeFigure} act is dedicated to looking at highlights, and therefore must have a corresponding highlight. The \textit{AskRelation} subact is special in that it needs to make two independently identifiable highlights, and for this reason, we provide a second brown pen for this subact. While we provide examples of how to highlight lines, angles, arcs, etc., in the annotator instructions, they are free to choose exactly how to proceed. An example highlight from our dataset is shown in \Cref{fig:high_ex}.

\subsection{\colorbox{fbcolor}{Feedback Selection}}\label{sec:fs}
Feedback, both positive \citep{martinez2024positive} and negative \citep{fossati2008role} is an important component in tutoring. However, the feedback component can often be small and inactionable, so making it an independent act or subact might not be the best idea. 
Therefore, we instead choose to give an option to add a positive or negative feedback to the start of the utterance by selecting the feedback type. There is also an option to give no feedback. The annotators indicate whether the student is on the right path or the wrong path, and the prompt is modified accordingly to provide positive or negative feedback. Note that pure feedback statements are still possible through the \textit{Continue} subact.


\section{Data Collection}
\subsection{Sourcing Problems and Student Errors}
\label{sec:PSE}
We draw our problems from the Geometry3k~ \citep{lu2021inter} and PGPS9k~ \citep{PGPS9k}. Geometry3k is a dataset of $3,000$ geometry problems with associated diagrams and annotations for diagram elements. PGPS9k is an open-source dataset which seeks to extend Geometry3k to $9,000$ problems. We make use of the entire training set of Geometry3k and $1,700$ additional problems from PGPS9k as our starting set. 


We then used \texttt{gemini-2.5-flash-preview-05-20} \citep{googleGeminiFlash}, hereafter referred to as Gemini, to identify and remove problems whose diagrams lacked point labels, since point labels provide an important mechanism for referring to specific diagram elements in instructional discourse. Although this filtering step may introduce some noise, it provides a practical way to reduce diagrams without usable point references.
Then, we used smaller models (details in App.\ref{eval_detail}) to generate potential student errors. For each question, we generated $10$ potential solutions, and only kept questions where at least two of these had the wrong final answer. Finally, we used Gemini to generate step-by-step correct solutions, dropping any questions whose final answer does not match the official answer. 
After applying all these filters, we were left with $1,422$ problems\footnote{The number of problems in the initial set was incrementally increased until we reached about $1,400$ problems, given our annotation budget.}.
\subsection{Data Annotation}
\label{sec:DA}
Our annotators were hired on Prolific\footnote{\url{https://www.prolific.com/}}. Qualifications were set to have math teaching experience and fluency in English\footnote{Follow-up interviews further supported participant suitability: all five interviewed participants provided responses consistent with our teaching-experience criteria, as discussed in \Cref{sec:int}. Together with the first-round screening, these interviews provide reasonable confidence in the validity of participants' reported qualifications.}. Participants first watched a 12-minute onboarding video explaining the annotation process and then answered a short quiz (see \Cref{fig:quiz}). Those who passed were invited to do one round of annotations. The authors evaluated the quality of the annotations and invited those with high-quality dialogs for further rounds. In the end, most of the dialogs were generated by 11 annotators\footnote{Ten additional annotators contributed a total of $89$ conversations between them. Most of them were not asked to return for further rounds due to poor quality dialogs}, 6 male, 5 female. Annotators were paid \pounds 2 for the onboarding task and \pounds14 per batch of $6$ dialogs. Furthermore, they received a bonus of \pounds0.10 for each diagram they highlighted, to encourage highlights and compensate for the additional effort. The entire process was approved by the ETH Ethics Committee.

\vspace{-2mm}
\subsection{Post Processing}

\label{sec:ann}
The free form highlights made by the annotators are a convenient way for humans to communicate, but looking at the highlights, it is easy to see that the true message being communicated has significantly fewer degrees of freedom. The vast majority of highlights can be expressed as either line segments, angles, arcs, and, at times, values marked within the diagram or point names. 
Moreover, the same target can be highlighted in multiple valid ways. For example, an angle may be indicated by circling it, tracing over its sides, or drawing an arc inside it, and even within each strategy, the exact highlight position may vary.
Since these differences usually carry no information, they are essentially noise, which would make both training and evaluation hard.

To deal with this issue, we standardise all highlights as lines, angles and arcs, identified by the named points forming them. The remaining highlights are consolidated into a fourth category: labels. These annotations are text-based and were produced by the first author.
Since these highlights are text based, they also open up the possibility of using VLMs that can't generate images. We discuss  how we can recover the highlights from these textual annotations in \Cref{sec:6.2}.


\section{Data Analysis}
\subsection{Data Statistics}

We collected $237$ dialog batches and, after filtering problematic cases (see \Cref{sec:filter} for details) retained $1,341$ dialogs. We split them into $800$ training, $141$ validation, and $400$ test examples.

The dialogs have about $8.6$ teacher utterances on average, with a median of $8$ and a maximum of $27$. About $45\%$ of these utterances have corresponding diagram highlights, with a median number of highlights at $3$ per conversation. For $84.6\%$ of these conversations, the annotators believe they led the student to the correct answer without revealing much of the answer, while another $13.6\%$ reached the final answer with substantial revealing. 

\begin{wrapfigure}{r}{0.5\textwidth}
\centering
\vspace{-20pt}
   \includegraphics[width=0.49\textwidth]{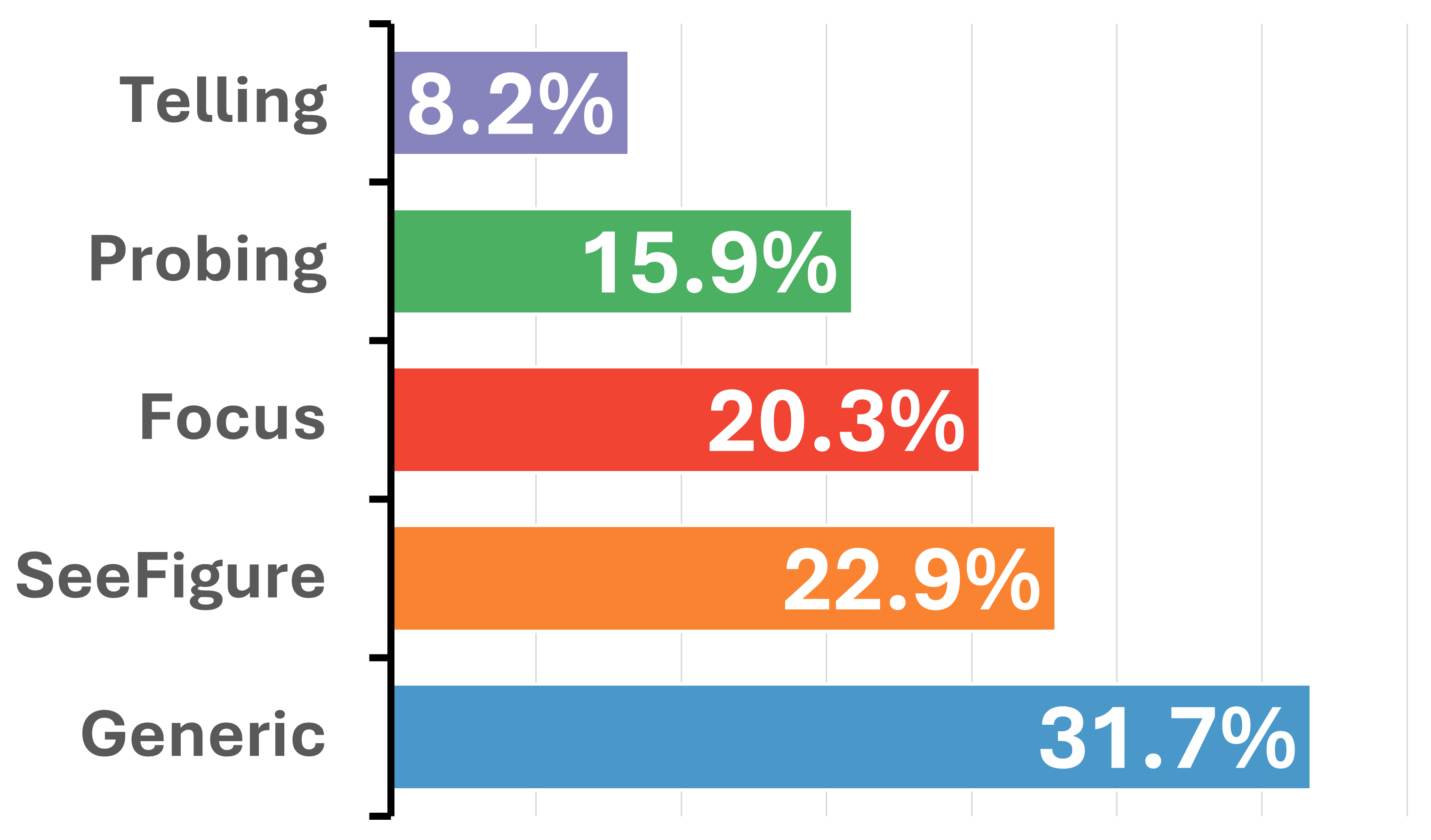}
    \caption{Dialog Act Statistics.}
    \vspace{-3mm}
    \label{fig:actper}
\end{wrapfigure}

All 3 subacts of \textit{\color{genericcolor}Generic} sit among the top-5 subacts with $11.3\%$, $11.2\%$ and $9.2\%$  respectively for \textit{Farewell}, \textit{Continue},  and \textit{Introduce}. \textit{Introduce} and \textit{Farewell} account for $78\%$ of first utterances and $90\%$ of final utterances respectively. Rounding up the top-5 subacts, we have \textit{Calculate} and \textit{GetRelation} at $9.9\%$ and $7.5\%$. A full pie chart of subacts is presented in \Cref{fig:subact_per}.
In the feedback option, positive feedback was selected in $53\%$ of the utterances, while negative feedback was selected in only $18\%$ of the cases. If we split every conversation into $4$ equal quarters\footnote{If a conversation has $n+1$ teacher utterances, utterance $i$ is in the $\lceil 4i/n\rceil^{th}$ quarter. The first utterance does not allow feedback and is hence left out of this analysis.}, we see that the fraction of positive feedback utterances grows from $40\%$ in the first quarter to $72\%$ in the final quarter, while the fraction of negative feedback falls from $33.5\%$ in the first quarter to only $6\%$ in the final quarter. This may be because student errors are more likely to occur earlier in the conversation than in later parts, when they have already been rectified.

For the utterances, Gemini's proposals were picked for $31.7\%$ of all utterances, the highest of all models. Proposals from Claude and GPT were picked in $25.6\%$ and $20.1\%$ of cases, respectively. In $22.5\%$ cases, annotators used the custom utterance option. While this includes cases where one of the proposals was copied and edited slightly, we note that for about $87\%$ of these utterances, the Levenshtein distance~ \citep{lcvenshtcin1966binary} to each of the 3 model proposals is $20\%$ or more than the total length of the utterance, indicating that these were more likely written from scratch.

\subsection{Student Simulation Fidelity}
\label{sec:fid}

Because students are role-played by VLMs, we evaluate whether the collected dialogs reflect realistic tutoring interactions. We use a two-pronged approach: after each conversation, annotators answer debrief questions about the fidelity of different components; we also interview 5 of the 11 main contributors to obtain a higher-level view of AI student behavior and teacher responses.

\subsubsection{Debrief Questions}
\label{sec:debrief}
After each conversation, we ask the annotators to rate Q1) Student Simulation Fidelity, Q2) Teacher Message Suggestion Fidelity, Q3) Usefulness of Teacher Message Suggestions, and Q4) Student's understanding of Diagrams, on a 5-point Likert scale. To reduce directional bias, the scales for Q1 and Q3 went low to high, while Q2 and Q4 went high to low. When reporting, we flip Q2 and Q4 so higher is better for all ratings.

 We find that all the 4 questions had a median score of $5$ while the 
 mean scores ranged from $4.28$ for Q3 to $4.52$ for Q1. 
 Overall, annotators judged the AI-simulated components to have high fidelity in most conversations.
 The full distributions for the ratings, and the full question texts can be found in \Cref{fig:debrief}.

\subsubsection{Annotator Interviews}\label{sec:int}
We interviewed 5 of our 11 main contributors to get a high level view of how realistic the AI students and their confusions felt, how it compared to their real life students, and most importantly, if the student fidelity was good enough such that the collected conversations would reflect their real teaching strategies. All of these teachers had at least three years of teaching experience and had interacted with at least 100 unique students.

A detailed interview analysis is provided in \Cref{sec:inter}. 
In summary, the interviewees found that \GD\ students were slightly more engaged than real students, but showed less emotional engagement, consistent with \citet{martynova-etal-2025-llms}. 
In diagram understanding and teacher frustration, \GD{} students were perceived as comparable to human students. Interviewees also reported that initial student confusions were largely indistinguishable from those of real students, and that they only slightly adapted their teaching strategies in \GD{}, suggesting that the dataset captures realistic tutoring behavior.


\subsection{Diagram Highlights}
In total, $5,168$ highlighted diagrams were annotated. Of these, $1,045$ included highlights made with the brown pen, including seven cases that did not use the green pen, contrary to the annotation instructions. Lines and angles were the most frequently highlighted elements, appearing in $2,644$ and $2,001$ diagrams, respectively. In addition, $799$ diagrams contained labels and $399$ included arcs. This standardization process necessarily abstracts over some visual nuance, especially for the broader `Labels' category. However, such cases account for less than $1\%$ of the annotations; we discuss them in detail in \Cref{sec:loss}.


\section{Experiments}
\label{sec:7}

Having collected and analysed \GD, we now proceed to finetune VLMs on it to generate teacher utterances.
We discuss how to convert these labels back into image highlights in  \Cref{sec:6.2}.

\subsection{Main Experiment}
\label{sec:5.1}
\textbf{Setup:} 
We emulate the annotation setup as closely as possible to preserve the original conditional dependencies and avoid circular prediction.
The model is given the context, comprising the question, the diagram (for models which can handle diagrams), the correct answer, the student answer and the conversation so far. Using these, the model must generate, in order, the act, the subact, the highlights, the feedback type and finally the teacher utterance. The prompt explains the setup, followed by brief descriptions of the acts, subacts, and feedback types. It also explains the output format, including how to output highlights in text. The full prompt including the output format is presented in \Cref{app:tp}). 

\noindent\textbf{Metrics:} We propose several metrics to evaluate tutor models.
As act, subact, and feedback are categorical, we report Macro-F1. For the final teacher utterance, we report BLEU \citep{10.3115/1073083.1073135} and BERTScore \citep{zhang2020bertscoreevaluatingtextgeneration}. For highlights, the output for each category is a set, so we could calculate the F1 score for each output and report the average over all instances. 
However, this F1 score is undefined when the ground-truth highlight set is empty, which is rather common in our task. 
Thus, we report 2 scores: \Pd\ and \D. \Pd\ Accuracy (and \Pd\ F1) looks at whether or not the model correctly predicted that it must make a highlight of the current category. It ignores whether the highlight is correct; e.g., if the ground truth is that line segment AB should be highlighted, the model is considered correct (true positive) as long as it highlights at least one line segment. 
If, however, the ground truth has no highlighted line segments, the model also must highlight no line segments to be considered correct (true negative). \D\ F1, on the other hand, is the average F1 score calculated over the instances where it is defined, i.e., where the ground truth is non-empty. We ignore highlight colors, merging green and brown highlights of the same type.

\noindent\textbf{Models:} We evaluate on 3 closed source models (Gemini, GPT and Claude, same ones used to generate teacher utterances), 3 open source text-only models (Qwen3 8B and 32B, Llama 3.1 8B) and 4 VLMs (Qwen3-VL 8B and 32B, InternVL 8B and Llama 3.2 11B Vision). For the open models, we evaluate them both with and without fine-tuning on \GD. We also report a \textit{Random Turn} baseline, which picks a random entry from the dataset as the output. This is a slightly better random baseline, as the outputs match the distribution of the dataset.

\begin{table*}[t]
\centering
\scriptsize
\setlength{\tabcolsep}{2.3pt}
\renewcommand{\arraystretch}{0.92}
\caption{Performance of models on the \GD\ test set. Text-only models are in \textit{{\colorbox[HTML]{E4DFF4}{lavender italics}}}. Best-performing models are in bold. See \Cref{tab:results_test_ci1,tab:results_test_ci2} for confidence intervals.}
\label{tab:results}
\resizebox{\textwidth}{!}{
\begin{tabular}{clccc|cccc|cccc|cc}
\toprule
\textbf{Group} &
\textbf{Model} &
\multicolumn{3}{c|}{\textbf{Categorical Macro-F1}} &
\multicolumn{4}{c|}{\textbf{Highlight Prediction Accuracy/F1}} &
\multicolumn{4}{c|}{\textbf{Decision F1}} &
\multicolumn{2}{c}{\textbf{Utterance}} \\
\cmidrule(lr){3-5}
\cmidrule(lr){6-9}
\cmidrule(lr){10-13}
\cmidrule(lr){14-15}
& &
\textbf{Act} & \textbf{Subact} & \textbf{Feedback}
& \textbf{Line} & \textbf{Angle} & \textbf{Arc} & \textbf{Label}
& \textbf{Line} & \textbf{Angle} & \textbf{Arc} & \textbf{Label}
& \textbf{BERT} & \textbf{BLEU} \\
\midrule

\multirow{1}{*}{\rotatebox[origin=c]{90}{}}
& Random Turn
& 25.5 & 12.8 & 47.4 & 68.2/23.5 & 71.6/18.7 & 93.8/5.5 & 86.2/9.1 & 0.2 & 0.3 & 0.0 & 0.2 & 15.3 & 3.4\\

\midrule

\multirow{3}{*}{\rotatebox[origin=c]{90}{\textbf{Closed}}}
& Gemini 2.5 Flash
& 30.4 & 17.5 & 55.6 & 63.4/46.2 & 72.7/44.0 & 94.0/37.5 & 82.9/17.4 & 18.1 & 10.3 & 1.7 & 4.3 & 25.2 & 5.4\\

& GPT-4o
& 30.0 & 13.6 & 51.4 & 39.9/38.3 & 72.3/37.0 & 94.6/33.0 & 91.4/3.4 & 20.1 & 6.7 & 0.9 & 0.1 & 21.5 & 3.6 \\

& Claude Sonnet 4
& 32.7 & 18.0 & 53.6 & 71.7/41.3 & 79.4/36.1 & 95.9/42.5 & 83.7/\textbf{19.5} & 11.4 & 5.9 & 0.9 & 6.7 & 25.7 & 6.1\\

\midrule

\multirow{7}{*}{\rotatebox[origin=c]{90}{\textbf{Open}}}
& {\cellcolor[HTML]{E4DFF4}\textit{Qwen 3 8B}}
& {\cellcolor[HTML]{E4DFF4}\textit{27.1 }}&{\cellcolor[HTML]{E4DFF4}\textit{ 11.3 }}&{\cellcolor[HTML]{E4DFF4}\textit{ 43.8 }}&{\cellcolor[HTML]{E4DFF4}\textit{ 41.7/40.1 }}&{\cellcolor[HTML]{E4DFF4}\textit{ 54.6/40.7 }}&{\cellcolor[HTML]{E4DFF4}\textit{ 91.2/36.4 }}&{\cellcolor[HTML]{E4DFF4}\textit{ 77.5/11.8 }}&{\cellcolor[HTML]{E4DFF4}\textit{ 19.9 }}&{\cellcolor[HTML]{E4DFF4}\textit{ 10.9 }}&{\cellcolor[HTML]{E4DFF4}\textit{ 0.0 }}&{\cellcolor[HTML]{E4DFF4}\textit{ 1.7 }}&{\cellcolor[HTML]{E4DFF4}\textit{ 17.1 }}&{\cellcolor[HTML]{E4DFF4}\textit{ 2.1}} \\

& {\cellcolor[HTML]{E4DFF4}\textit{Llama 3.1 8B}}
& {\cellcolor[HTML]{E4DFF4}\textit{17.1 }}&{\cellcolor[HTML]{E4DFF4}\textit{ 5.4 }}&{\cellcolor[HTML]{E4DFF4}\textit{ 45.5 }}&{\cellcolor[HTML]{E4DFF4}\textit{ 39.6/38.3 }}&{\cellcolor[HTML]{E4DFF4}\textit{ 75.1/28.2 }}&{\cellcolor[HTML]{E4DFF4}\textit{ 95.2/29.2 }}&{\cellcolor[HTML]{E4DFF4}\textit{ 86.4/9.2 }}&{\cellcolor[HTML]{E4DFF4}\textit{ 14.3 }}&{\cellcolor[HTML]{E4DFF4}\textit{ 2.2 }}&{\cellcolor[HTML]{E4DFF4}\textit{ 0.4 }}&{\cellcolor[HTML]{E4DFF4}\textit{ 0.4 }}&{\cellcolor[HTML]{E4DFF4}\textit{ 19.0 }}&{\cellcolor[HTML]{E4DFF4}\textit{ 3.0}} \\

& {\cellcolor[HTML]{E4DFF4}\textit{Qwen 3 32B}}
& {\cellcolor[HTML]{E4DFF4}\textit{29.5 }}&{\cellcolor[HTML]{E4DFF4}\textit{ 15.1 }}&{\cellcolor[HTML]{E4DFF4}\textit{ 51.2 }}&{\cellcolor[HTML]{E4DFF4}\textit{ 46.6/41.4 }}&{\cellcolor[HTML]{E4DFF4}\textit{ 66.0/45.3 }}&{\cellcolor[HTML]{E4DFF4}\textit{ 92.2/41.1 }}&{\cellcolor[HTML]{E4DFF4}\textit{ 86.5/9.3 }}&{\cellcolor[HTML]{E4DFF4}\textit{ 19.2 }}&{\cellcolor[HTML]{E4DFF4}\textit{ 11.0 }}&{\cellcolor[HTML]{E4DFF4}\textit{ 0.0 }}&{\cellcolor[HTML]{E4DFF4}\textit{ 1.7 }}&{\cellcolor[HTML]{E4DFF4}\textit{ 20.3 }}&{\cellcolor[HTML]{E4DFF4}\textit{ 3.1}}\\

& Qwen 3 VL 8B
& 30.4 & 13.7 & 54.9 & 41.9/39.2 & 66.4/42.9 & 95.1/27.9 & 54.0/13.8 & 16.6 & 10.0 & 1.3 & 4.8 & 21.1 & 3.6 \\

& InternVL 8B
& 3.6 & 1.0 & 13.7 & 71.0/21.4 & 74.8/13.6 & 96.2/3.0 &\textbf{ 92.1}/0.8 & 3.3 & 1.0 & 0.0 & 0.1 & 3.5 & 0.6\\

& Llama 3.2 11B-V
& 19.4 & 6.0 & 44.3 & 40.6/38.0 & 75.1/32.1 & 95.4/31.1 & 87.6/10.0 & 13.5 & 2.4 & 0.4 & 0.6 & 18.9 & 2.9\\

& Qwen 3 VL 32B
& 29.4 & 12.9 & 53.4 & 42.2/40.2 & 52.5/41.4 & 94.1/41.1 & 59.9/17.8 & \textbf{20.2} &\textbf{ 13.4} & 1.7 & \textbf{10.0} & 21.3 & 4.2 \\

\midrule

\multirow{7}{*}{\rotatebox[origin=c]{90}{\textbf{Fine-tuned}}}
& {\cellcolor[HTML]{E4DFF4}\textit{Qwen 3 8B}}
& {\cellcolor[HTML]{E4DFF4}\textit{45.4 }}&{\cellcolor[HTML]{E4DFF4}\textit{ 28.6 }}&{\cellcolor[HTML]{E4DFF4}\textit{ 71.6 }}&{\cellcolor[HTML]{E4DFF4}\textit{ 79.3/44.0 }}&{\cellcolor[HTML]{E4DFF4}\textit{ 83.7/47.8 }}&{\cellcolor[HTML]{E4DFF4}\textit{ 96.4/35.3 }}&{\cellcolor[HTML]{E4DFF4}\textit{ 90.0/19.3 }}&{\cellcolor[HTML]{E4DFF4}\textit{ 8.9 }}&{\cellcolor[HTML]{E4DFF4}\textit{ 5.6 }}&{\cellcolor[HTML]{E4DFF4}\textit{ 1.5 }}&{\cellcolor[HTML]{E4DFF4}\textit{ 3.2 }}&{\cellcolor[HTML]{E4DFF4}\textit{ 34.0 }}&{\cellcolor[HTML]{E4DFF4}\textit{ 13.0}} \\

& {\cellcolor[HTML]{E4DFF4}\textit{Llama 3.1 8B}}
& {\cellcolor[HTML]{E4DFF4}\textit{43.7 }}&{\cellcolor[HTML]{E4DFF4}\textit{ 28.8 }}&{\cellcolor[HTML]{E4DFF4}\textit{ 69.0 }}&{\cellcolor[HTML]{E4DFF4}\textit{ \textbf{80.7}/45.2 }}&{\cellcolor[HTML]{E4DFF4}\textit{ 83.6/44.8 }}&{\cellcolor[HTML]{E4DFF4}\textit{ 96.6/37.1 }}&{\cellcolor[HTML]{E4DFF4}\textit{ 90.8/18.9 }}&{\cellcolor[HTML]{E4DFF4}\textit{ 9.1 }}&{\cellcolor[HTML]{E4DFF4}\textit{ 5.5 }}&{\cellcolor[HTML]{E4DFF4}\textit{ 4.0 }}&{\cellcolor[HTML]{E4DFF4}\textit{ 2.6 }}&{\cellcolor[HTML]{E4DFF4}\textit{ 33.1 }}&{\cellcolor[HTML]{E4DFF4}\textit{ 12.7}} \\

& {\cellcolor[HTML]{E4DFF4}\textit{Qwen 3 32B}}
& {\cellcolor[HTML]{E4DFF4}\textit{45.6 }}&{\cellcolor[HTML]{E4DFF4}\textit{ 28.9 }}&{\cellcolor[HTML]{E4DFF4}\textit{ 71.8 }}&{\cellcolor[HTML]{E4DFF4}\textit{ 80.3/46.0 }}&{\cellcolor[HTML]{E4DFF4}\textit{ 82.7/44.4 }}&{\cellcolor[HTML]{E4DFF4}\textit{ 96.6/42.4 }}&{\cellcolor[HTML]{E4DFF4}\textit{ 90.3/11.9 }}&{\cellcolor[HTML]{E4DFF4}\textit{ 9.7 }}&{\cellcolor[HTML]{E4DFF4}\textit{ 6.3 }}&{\cellcolor[HTML]{E4DFF4}\textit{ 2.3 }}&{\cellcolor[HTML]{E4DFF4}\textit{ 1.9 }}&{\cellcolor[HTML]{E4DFF4}\textit{ 35.0 }}&{\cellcolor[HTML]{E4DFF4}\textit{ 14.1}}\\

& Qwen 3 VL 8B
& 46.4 & 30.9 & 71.7 & 79.9/45.5 & 84.0/\textbf{49.0} & 96.5/37.2 & 90.3/16.5 & 9.5 & 8.6 & 3.5 & 2.9 & 34.5 & 13.8 \\

& InternVL 8B
& 44.0 & 29.2 & 70.3 & 79.8/45.4 & 82.5/44.8 & 96.1/37.1 & 90.1/14.5 & 9.8 & 6.8 & 5.3 & 2.5 & 33.8 & 12.9 \\

& Llama 3.2 11B-V
& 43.5 & 27.8 & 69.1 & 79.9/42.9 & 83.4/45.4 & 96.7/40.6 & 90.2/11.9 & 9.0 & 5.6 & 3.6 & 2.2 & 33.1 & 12.7\\

& Qwen 3 VL 32B
& \textbf{46.5} & \textbf{31.2} & \textbf{72.1} & 80.4/\textbf{47.9} & \textbf{84.3}/47.0 & \textbf{96.7}/\textbf{45.9} & 90.5/15.5 & 10.9 & 7.2 & \textbf{7.3} & 3.1 & \textbf{35.4} & \textbf{14.3} \\
\bottomrule
\end{tabular}
}
\end{table*}

\noindent\textbf{Results:} \Cref{tab:results} reports test-set results, with validation set results in \Cref{tab:results_val}. 
For closed models and open-source models without fine-tuning, outputs sometimes violate the required format and cannot be parsed; in such cases, we assign a score of 0 for categorical variables and teacher utterances, and treat highlight outputs as empty sets.

\textbf{Categorical variables and teacher utterances follow the expected trends.} Closed models outperform open-source models without fine-tuning, while fine-tuned models perform best overall. 
A qualitative analysis comparing utterances from the fine-tuned models with those from the base models shows that fine-tuning improves scaffolding, dialog-flow control, and grounding in the highlighted regions (see \Cref{sec:qual-case-study} for details).
Vision-enabled models generally outperform text-only models, although the gap is sometimes small.
Larger models tend to perform better, with Qwen3-VL-32B achieving the strongest overall results.
Most models outperform the Random Turn baseline.

\textbf{Highlight prediction and selection remain challenging.}
Without fine-tuning, lines and angles achieve non-trivial \Pd\ and \D\ scores, but performance remains weak. Arcs and labels obtain high \Pd\ accuracy because they are rare and empty-set predictions are often correct, yet their F1 scores remain low. 
InternVL illustrates this pattern: formatting errors often result in empty-set predictions, inflating Prediction accuracy for rare categories.
As a result, the \D{} scores for these categories are negligible.


\textbf{Fine-tuning improves Prediction but can hurt Decision.}
Fine-tuning improves Prediction across the board, but lowers Decision scores for lines and angles. 
A zero Decision F1 can result either from predicting highlights that miss all correct ones, or from leaving the instance empty. 
As shown in \Cref{tab:missleave}, fine-tuning substantially decreases misses, but increases leaves enough to reduce the overall Decision scores. 
Thus, fine-tuned models learn to abstain from highlight selection more often, which is unsurprising given the strong empty-set imbalance in the data: the empty set is correct for approximately 77\% of line instances and 83\% of angle instances, with even higher rates for arcs and labels. 
This imbalance also helps explain their weak Decision scores. 
Finally, because Decision F1 does not penalize models for predicting highlights when none are needed, it should be interpreted together with Highlight F1, which captures such false-positive cases.


\subsection{On the Challenge of Highlight Generation}
\label{sec:5.3}

\begin{table*}[t]
\centering
\footnotesize
\setlength{\tabcolsep}{2.4pt}
\renewcommand{\arraystretch}{0.95}
\caption{Results of focused experiments on highlighting. ``From Model'' indicates that provided teacher utterances come from a Qwen-3-VL-8b model trained to generate teacher utterances without seeing the ground truth highlights. Combined Scores are calculated in accordance with \Cref{sec:combined}. See \Cref{tab:focus_c1} for confidence intervals.}
\label{tab:focus}

\newcolumntype{C}[1]{>{\centering\arraybackslash}p{#1}}

\begin{tabular}{
C{0.9cm}
C{1.55cm}
C{1.05cm}
@{\hspace{4pt}}|@{\hspace{4pt}}
cccc
@{\hspace{4pt}}|@{\hspace{4pt}}
ccc
@{\hspace{4pt}}|@{\hspace{4pt}}
ccc
}
\toprule
\textbf{Element} &
\textbf{Teacher} &
\textbf{Model} &
\multicolumn{4}{c@{\hspace{4pt}}|@{\hspace{4pt}}}{\textbf{Prediction (\%)}} &
\multicolumn{3}{c@{\hspace{4pt}}|@{\hspace{4pt}}}{\textbf{Decision, Indep. (\%)}} &
\multicolumn{3}{c}{\textbf{Combined Score (\%)}} \\
\cmidrule(lr){4-7}
\cmidrule(lr){8-10}
\cmidrule(lr){11-13}
& & &
\textbf{Prec.} & \textbf{Rec.} & \textbf{F1} & \textbf{Acc.} &
\textbf{Prec.} & \textbf{Rec.} & \textbf{F1} &
\textbf{Prec.} & \textbf{Rec.} & \textbf{F1} \\
\midrule

\multirow{7}{*}{\textbf{Line}}
& \multirow{3}{*}{From Model}
& Weighted
& 74.8 & 56.8 & 64.5 & 84.7
& \multicolumn{3}{c@{\hspace{4pt}}|@{\hspace{4pt}}}{Same as fine-tuned model}
& 47.4 & 36.0 & \textbf{40.9} \\
& & Yes
& 67.9 & 55.5 & 61.1 & 82.6
& 63.1 & 62.3 & 59.9
& 43.7 & 35.8 & 39.3\\
& & No
& 38.8 & 85.4 & 53.3 & 63.3
& 61.1 & 58.3 & 57.2
& 22.2 & 48.8 & 30.5 \\
\cmidrule(lr){2-13}

& \multirow{2}{*}{Ground Truth}
& Yes
& 77.7 & 65.6 & 71.1 & 86.9
& 78.1 & 78.8 & 76.8
& 66.3 & 56.0 & 60.7 \\
& & No
& 39.0 & 84.7 & 53.4 & 63.6
& 78.2 & 73.9 & 74.1
& 29.6 & 64.3 & 40.5 \\
\cmidrule(lr){2-13}

& \multirow{2}{*}{Not Given}
& Yes
& 69.7 & 55.2 & 61.6 & 83.1
& 62.9 & 64.8 & 61.3
& 44.3 & 35.1 & 39.2 \\
& & No
& 39.9 & 83.1 & 54.0 & 65.1
& 51.5 & 44.0 & 44.3
& 17.3 & 36.0 & 23.4 \\

\midrule

\multirow{7}{*}{\textbf{Angle}}
& \multirow{3}{*}{From Model}
& Weighted
& 61.6 & 51.8 & 56.3 & 85.9
& \multicolumn{3}{c@{\hspace{4pt}}|@{\hspace{4pt}}}{Same as fine-tuned model}
& 29.6 & 24.9 & \textbf{27.1} \\
& & Yes
& 59.7 & 55.9 & 57.7 & 85.6
& 48.3 & 47.6 & 46.9
& 27.5 & 25.8 & 26.6 \\
& & No
& 32.5 & 81.8 & 46.6 & 67.0
& 43.9 & 44.2 & 42.9
& 13.6 & 34.2 & 19.5 \\
\cmidrule(lr){2-13}

& \multirow{2}{*}{Ground Truth}
& Yes
& 74.5 & 62.3 & 67.8 & 89.6
& 61.6 & 61.3 & 60.7
& 48.4 & 40.5 & 44.1 \\
& & No
& 36.1 & 87.7 & 51.1 & 70.5
& 57.8 & 56.9 & 56.6
& 20.6 & 50.1 & 29.2 \\
\cmidrule(lr){2-13}

& \multirow{2}{*}{Not Given}
& Yes
& 61.4 & 56.4 & 58.8 & 86.1
& 41.6 & 42.1 & 40.8
& 26.2 & 24.1 & 25.1 \\
& & No
& 33.8 & 79.5 & 47.5 & 69.1
& 44.2 & 42.3 & 41.6
& 13.4 & 31.5 & 18.8 \\

\bottomrule
\end{tabular}
\vspace{-5mm}
\end{table*}

\textbf{Focused Highlight Model:} In \Cref{sec:5.1}, teacher utterances and diagram highlights are generated together in a structured format using a single VLM call. While convenient, this requires the model to learn multiple subtasks simultaneously, including highlight prediction and output formatting.
Since highlights appear particularly challenging, we investigate whether a more focused setup improves highlight generation. We focus on the two most common highlight types in our dataset, lines and angles, and split the task of \D\ and \Pd\ into two separate models. This results in four tasks: \D\ and \Pd\ for lines and angles. The model input consists of the dialogue context, the act and the subact. The \D\ model is only called when the ground truth highlights are non-empty, and is explicitly instructed to output at least one highlight. All experiments use Qwen3-VL-8B and the validation set.

The results of focused experiments are shown in \Cref{tab:focus}. The base \D\ F1, although substantially higher than in the previous setting, remains relatively low. 
Since the highlight model is now separated from utterance generation, we can reverse their generation order.
We observe that highlight performance improves significantly when highlights are generated using ground-truth utterances as input. This behavior is expected, as the ground truth utterances were originally generated \textit{after} the teacher made highlights, and can often make direct reference to what was highlighted. But this same fact makes this setting unrealizable at inference. To avoid this, we instead train a separate utterance generation model and use the utterances from this model instead. The utterances from this model do not have access to the highlights, and thus, even though it performs worse than the ground truth version, it is realistic.

\noindent\textbf{Weighted training:}
Next, we explore alternate training strategies for highlight generation.
In \Cref{sec:5.1}, we observed that training the model causes it to produce fewer highlights, potentially due to unbalanced data. To test this hypothesis, we adopted a weighted training approach for the highlight model, where each positive sample was repeated three times per epoch, and each negative sample appeared once per epoch.
This generally improved highlighting performance (see Weighted in \Cref{tab:focus}).

\noindent\textbf{Combined Metric:} Finally, to directly compare settings above, we define a combined F1 score. 
The precise formulation is described in \Cref{sec:combined}.
In brief, this metric is similar to the \Pd\ F1 score, except that each True Positive contributes its corresponding \D\ F1 score instead of a value of 1.
The Combined Recall corresponds to the \D\ F1 reported in \Cref{tab:results}. 
 Looking at the Combined scores, we observe that the Combined Recall follows a similar trend to the results in \Cref{tab:results}, with untrained models outperforming finetuned ones. However, the Combined F1 scores indicate that training is beneficial overall. 
 The best-performing viable models are the ones using {\bf weighted training}, for both lines and angles.



\subsection{Ablations}
To better understand the task, we run two sets of ablations focusing on the categorical variables and the Teacher utterance. 
We use Qwen3-VL-8B throughout, except for ablations without diagrams, where we use Qwen3-8B.

\noindent\textbf{Data Size:} We set out to collect approximately $1,400$ dialogs and ended up with $1,341$ after filtering. 
While more training data is usually helpful, there are often diminishing returns \citep{kaplan2020scalinglawsneurallanguage}. To understand how model performance scales with data size, we finetune our model by increasing the training set size from $0$ to $800$ in increments of $200$. The results are recorded in \Cref{tab:data_size}. We note that the biggest improvement comes between $0$ and $200$, proving that even a little data can be quite useful. By the time we reach the full training set, act and subact scores have saturated, though we still see improvements on Feedback and teacher utterance. While the latter indicate that some more data could still be helpful, overall, the training data is sufficiently large.

%

\begin{wraptable}{r}{0.64\textwidth}
\vspace{-8pt}
\centering
\small
\setlength{\tabcolsep}{2.2pt}
\caption{\textbf{Context ablation analysis.}
OLS coefficients for each context component before and after fine-tuning. }
\label{tab:abl2}
\begin{tabular}{l|cc|cc|cc|cc}
\toprule
\textbf{Context} &
\multicolumn{2}{c|}{\textbf{Act}} &
\multicolumn{2}{c|}{\textbf{Subact}} &
\multicolumn{2}{c|}{\textbf{Feedback}} &
\multicolumn{2}{c}{\textbf{BLEU}} \\
\cmidrule(lr){2-3}
\cmidrule(lr){4-5}
\cmidrule(lr){6-7}
\cmidrule(lr){8-9}
& \textbf{FT} & \textbf{Base}
& \textbf{FT} & \textbf{Base}
& \textbf{FT} & \textbf{Base}
& \textbf{FT} & \textbf{Base} \\
\midrule
Question
& 0.29 & 0.14
& -0.10 & -0.66
& 0.08 & -0.35
& 0.29 & -0.19 \\
Diagram
& 0.87 & -0.31
& 0.73 & -1.52
& 0.44 & 2.15
& -0.01 & -0.06 \\
Correct Sol.
& 0.71 & 1.10
& 1.46 & 0.60
& 3.73 & 7.14
& 0.47 & 0.23 \\
Student Sol.
& 0.16 & -0.66
& -0.68 & -0.20
& 0.13 & -4.01
& 0.04 & -0.19 \\
Constant
& 42.75 & 32.01
& 28.30 & 14.62
& 66.61 & 48.57
& 11.63 & 3.42 \\
$R^2$
& 28.2 & 38.0
& 43.7 & 78.8
& 82.3 & 79.9
& 79.8 & 72.1 \\
\bottomrule
\end{tabular}
\vspace{-8pt}
\end{wraptable}
\noindent\textbf{Which Context Elements Matter?}
Our models generate the tutor response and highlights conditioned on 
the question text, the diagram, the correct solution, the student solution and the conversation so far. 
A natural question that arises is how much each of these components contributes to the modeling. Of these, the role of the conversation is clear. In order to generate the next conversational turn, the model must know where in the conversation it is. In this experiment, we try to understand the importance of the other four elements, by considering all $2^4=16$ combinations, both with and without model training. The full set of results is reported in \Cref{tab:abl_full}.

To get a better view of the importance of each element, we run separate Ordinary Least Squares (OLS) regressions for finetuned and unfinetuned models, the results of which are shown in \Cref{tab:abl2}. 
We notice that without finetuning, extra elements in the context almost always cause a decline in performance. The only exception to this is the Correct Solution, which always helps. With finetuning, the correct solution continues to have the highest coefficients, though now most coefficients are positive, indicating that the model learns to make use of the other components.  

\section{Related Work}
Prior research relevant to \GD\ spans three areas: 
(1) Mathematics education studies highlight the importance of visual representations and cues for learning \citep{kaitera2022developing, boonen2016s, small2025eyes, MAYER200285, evagorou2015role, cooper2018benefits, arcavi2003role, SCHOENHERR2024100639, jones2013diagrams};
(2) Prior educational dialog datasets are largely text-only and do not capture diagram-grounded interactions \citep{macina2023mathdial, stasaski-etal-2020-cima, suresh2022fine, demszky-hill-2023-ncte, caines2020teacher, wang-etal-2024-book2dial}; (3) Previous multimodal geometry datasets have advanced diagram understanding and geometry problem solving. However, they mainly focus on solving problems rather than modeling how teachers use diagrams in instruction \citep{Seo_Hajishirzi_Farhadi_Etzioni_2014, seo-etal-2015-solving, lu2021inter, PGPS9k, chen-etal-2021-geoqa, chen-etal-2022-unigeo, ICLR2025_09afabe3, xu2025geosense}. \GD\ connects these strands by introducing multimodal tutoring dialogs grounded in geometry diagrams, together with fine-grained visual highlight annotations, linking instructional language to pedagogically meaningful visual references. 
See Appendix~\ref{sec:related_work} for detailed related work.

\section{Conclusion}

We introduce \GD, to our knowledge the first dataset of geometry tutoring dialogs that tightly integrates conversational teaching strategies with diagram-based visual guidance.
\GD\ captures rich pedagogical behavior, with nearly half of tutoring turns grounded in diagram highlights and a diverse distribution of instructional strategies.
Through fine-tuning experiments, we show that current VLMs can learn to generate higher-quality tutor utterances, but still struggle to produce meaningful diagram highlights, even with specialized training.
This gap highlights visually grounded pedagogical action as a distinct challenge for multimodal tutoring, and positions \GD\ as a resource for developing models that better coordinate instructional language with visual guidance.
\section*{Ethics Statement}
This study includes research with human subjects. This was approved by the ETH Ethics Committee as project 25-ETHICS-194. All interaction with human participants was made via Prolific, who ensure the anonymity of their participants. 
\section*{Acknowledgments and Disclosure of Funding}
This research was supported by the ETH AI Center through a doctoral fellowship to Junling Wang, the Swiss AI Large Grant SCR1089274, and Swiss AI Small Grant \#63. Sankalan Pal Chowdhury's Doctoral Fellowship is partially suported by the ETH-EPFL Joint Doctoral Program in Learning Science. Mrinmaya Sachan is supported by SNSF Grant number 10009282. Additionally, the authors thank all participants for their contribution to the dataset creation.

\bibliographystyle{plainnat}
\bibliography{custom}

\appendix
\section{Limitations and Societal Impact}
\label{app:lim}
In this paper, we propose a novel dataset exploring dialog based tutoring in textual and visual modalities. While \GD\ does make use of visuals, it is still rather limited in the type of visuals. We instruct our annotators to highlight parts of an existing diagram, which leaves out several other visual interactions, even within the limited purview of geometry, like constructions, spontaneous sketching, etc. Furthermore, these interactions are one-way, i.e., the student cannot modify the diagram. It is also worth noting that drawing with a mouse is difficult, which likely discouraged annotators from making complex highlights, something that would not have occurred if they had used pen and paper or a blackboard. 

Further, we must also acknowledge that Geometry forms only a small part of mathematics and an even smaller part of education in general. Tutoring techniques vary widely across these subjects, so more datasets need to be collected for other subjects. Visual interactions in other domains, such as Physics may require significant adaptations such as drawing free-body diagrams, etc. which we leave for future work.

Another shortcoming is the fact that the ground truth teacher utterance and highlights are one of potentially more than one equally valid next step at any given point. This limitation, however, is not unique to us, and plagues most work in Natural Language Processing, as there is typically more than one correct way to express something.

Finally, in our approach, both the student errors and student utterances are generated by VLMs, which might not accurately represent real students. We attempt to address this issue by our debrief questions, and interviews, which yield highly positive responses. However, we must note that teachers are not trained to detect fake students, so their judgment is not necessarily absolute. Even if they are right and the models simulate a single student quite well, there is no reason to believe that all types of students are equally represented. Since our dataset is aimed at training teacher models, a reasonable but not perfect student might serve our purposes. However, \textbf{
\GD\ dialogs should not be used to train student models.}

\GD\ is meant to be used to make LLMs better at visual tutoring, a task which has traditionally been the domain of human tutors. While this may lead to concerns about the potential replacement of humans with AI, it is far more likely to slot in to areas where human alternatives do not exist, due to the shortage of good teachers. This is likely to help students significantly. Such methods can also make the job of existing teachers easier by taking over more repetitive tasks.
\section{Issues with Standardising Highlights}
\label{sec:loss}
\begin{figure}
\centering
\begin{subfigure}{.5\linewidth}
  \centering
  \includegraphics[width=.99\linewidth, trim={0.3cm 0.3cm 0.3cm 0}, clip]{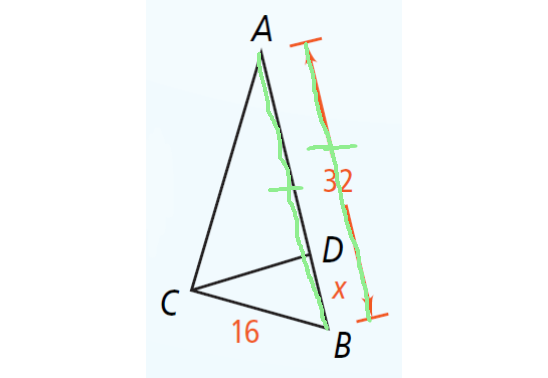}
  \label{fig:sub1}
\end{subfigure}%
\begin{subfigure}{.5\linewidth}
  \centering
  \includegraphics[width=.99\linewidth, trim={1.5cm 0.3cm 1.5cm 0}, clip]{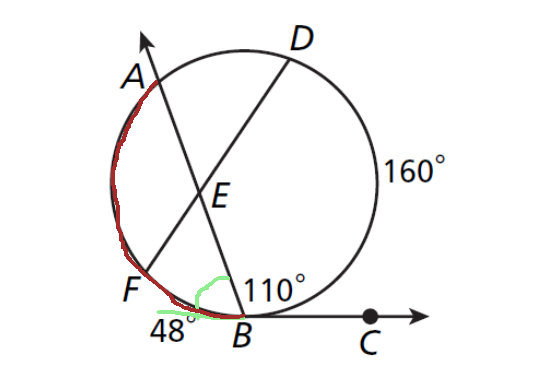}
  \label{fig:sub2}
\end{subfigure}
\caption{In the left image, the length marker, which is not part of the diagram is highlighted. On the right, The complement of $\angle$EBC is highlighted by extending BC. While both these are meaningful, our annotation schema does not accommodate them
}

\label{fig:high_diff}
\end{figure}

While this highlighting makes the task much more structured, we must note that it takes away some amount of nuance. As an example, annotators at times wrote out certain values on the diagram, which do not transfer to these annotations. We also found more creative use of highlights. For example, one annotator marked four different pairs of angles in four different ways to indicate that each pair was equal. While we were able to annotate which angles were highlighted, we could not differentiate between the different styles used. Two more such examples are shown in \Cref{fig:high_diff}. Some of the more common peculiar highlights were accounted for by using the ``Labels'' field as a miscellaneous entry. While these were not negligible, they only form a small fraction of all highlight annotations. In particular, of the $799$ turns had labels highlighted, for a total of $1172$ label highlights. Of these, $217$ were referring to points on the figure, $49$ were referring to parallel markers, $232$ were referring to equal markers and $651$ referred to text on the image. The Remaining $23$ accounted for all the miscellaneous categories, a mere $2.0\%$ of label highlights and $0.25\%$ of total highlights.

Despite our attempts to avoid this in the question setup, there were cases in which highlighted elements could not be annotated because the associated points were not named in the figure. Luckily, in all such cases, the required points had been named correctly in the data source (Geometry3K or PGPS9K), so we simply used these names and marked the points as ``Missing''. 
\begin{figure}
\centering
\begin{subfigure}{.5\linewidth}
  \centering
  \includegraphics[width=.99\linewidth, trim={0.3cm 0.3cm 0.3cm 0}, clip]{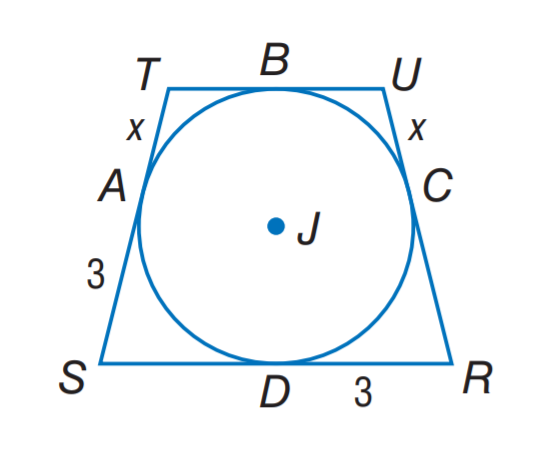}
  \caption{Original}
  \label{fig:sub1}
\end{subfigure}%
\begin{subfigure}{.5\linewidth}
  \centering
  \includegraphics[width=.99\linewidth, trim={1.5cm 0.3cm 1.5cm 0}, clip]{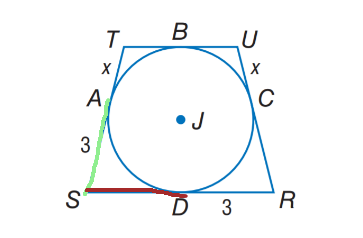}
  \caption{With Highlight}
  \label{fig:sub2}
\end{subfigure}
\caption{Example of a line highlight on a diagram. The Tutor is possibly trying to get the student to realise that $SA$ and $SD$ are equal.
}
\label{fig:high_ex}
\end{figure}
\section{Diagram Highlights}
\label{sec:auto_highlights}

\subsection{Example of Numeric Overlay}
\begin{figure}
    \centering
    \includegraphics[width=0.5\linewidth, trim={0.3cm 0 0 0.3cm}, clip]{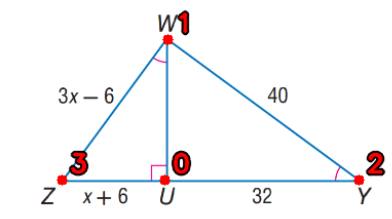}
    \caption{Example of the temporary numeric overlay on a diagram used for node-label matching.}
    \label{fig:numeric_overlay}
\end{figure}

We include an example of the temporary numeric overlay in Figure~\ref{fig:numeric_overlay}, showing how each detected point is assigned a red numeric marker within a black boundary.

\subsection{Examples of Automatic Diagram Highlights} \label{sec:example_highlight}
\begin{figure}
    \centering
    \includegraphics[width=0.5\linewidth, trim={0.3cm 0 0 0.3cm}, clip]{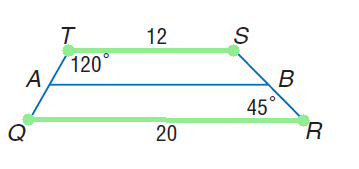}
    \caption{Example of automatic line highlighting.}
    \label{fig:line_example}
\end{figure}

\begin{figure}
    \centering
    \includegraphics[width=0.5\linewidth, trim={0.3cm 0 0 0.3cm}, clip]{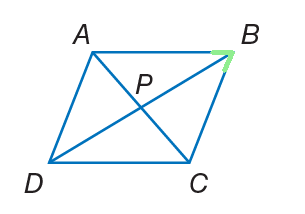}
    \caption{Example of automatic angle highlighting.}
    \label{fig:angle_example}
\end{figure}

\begin{figure}
    \centering
    \includegraphics[width=0.4\linewidth, trim={0.3cm 0 0 0.3cm}, clip]{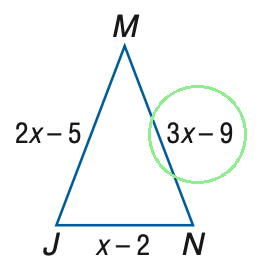}
    \caption{Example of automatic label highlighting.}
    \label{fig:label_example}
\end{figure}

\begin{figure}
    \centering
    \includegraphics[width=0.5\linewidth, trim={0.3cm 0 0 0.3cm}, clip]{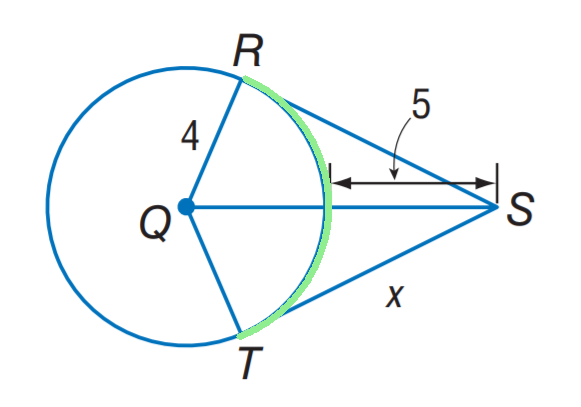}
    \caption{Example of automatic arc highlighting.}
    \label{fig:arc_example}
\end{figure}

We present representative examples of each supported highlight type generated by our pipeline. 
Figure~\ref{fig:line_example} illustrates automatic line highlighting, 
Figure~\ref{fig:angle_example} shows angle highlighting, 
Figure~\ref{fig:label_example} demonstrates label highlighting, 
and Figure~\ref{fig:arc_example} displays arc highlighting. 

\section{Acts and Subacts}
See table \ref{tab:subacts} for details of our subacts.
{
\hyphenpenalty=10000
\exhyphenpenalty=10000

\begin{table*}[t]
\centering
\scriptsize
\setlength{\tabcolsep}{4.0pt}
\renewcommand{\arraystretch}{1.08}
\caption{Full list of subacts used for \GD.}
\label{tab:subacts}

\newcolumntype{L}[1]{>{\raggedright\arraybackslash}p{#1}}

\begin{tabular}{
L{1.7cm}
L{2.5cm}
L{4cm}
L{4cm}
}
\toprule
\textbf{Act: Subact} &
\textbf{Description} &
\textbf{Usage Guidelines} &
\textbf{Examples} \\
\midrule

\textbf{Generic: Introduce}
&
Greet the student and ask them about their approach to solving the problem.
&
Only used as the first utterance of a conversation.
&
Can you explain how you decided to approach this problem? Just go through your thinking step by step. \\

\midrule

\textbf{Generic: Farewell}
&
Congratulate the student on solving the problem, and bid them goodbye.
&
Only used as the last utterance of the conversation.
&
That's right! Good job. Thanks for your time, and feel free to come back if you have more questions. Have a great day! \\

\midrule

\textbf{Generic: Continue}
&
Ask the student to continue talking about whatever they were saying.
&
Use when you want to skip your turn, e.g., when the student gave only part of the solution.
&
{\color[HTML]{B7B7B7}I see the right angle at M, $\triangle$LMN is a right triangle.}
That's a good observation. What does that tell you about the problem? \\

\midrule

\textbf{SeeFigure: AskLength}
&
Ask the student about the length of the highlighted arc or line segment.
&
\multirow{2}{4cm}{Use when you need the student to say the measure or length of an element. Usually, you want to take small steps, asking students to only state values that are either given already, or can be evaluated directly from given and previously stated values.}
&
Now, can you tell me the length of LP? \\

\cmidrule(lr){1-2}
\cmidrule(lr){4-4}

\textbf{SeeFigure: AskAngle}
&
Ask the student about the measure of the highlighted angle.
&
&
Can you calculate the measure of angle MLN? \\

\midrule

\textbf{SeeFigure: AskRelation}
&
Ask the student about the relation between two highlighted elements.
&
Two angles are equal, complementary, or supplementary; two line segments are equal; two triangles are congruent or similar; etc.
&
How do the angles marked in green and brown relate to each other? \\

\midrule

\textbf{SeeFigure: SeeHighlight}
&
Ask the student to look at a particular part of the figure.
&
Use when you need to consult the figure, but none of the previous three substrategies apply.
&
Look at the circled part of the figure. What can you say about line $l$? \\

\midrule

\textbf{Focus: SeekStrategy}
&
Ask the student what they should do next.
&
The student is doing well, so ask them to keep going. This is often interchangeable with Continue. Prefer SeekStrategy when the previous step is complete and the student must start a new step.
&
What should the next step be? \\

\midrule

\textbf{Focus: Guide}
&
Give the student a hint about what to do next.
&
Use when SeekStrategy fails.
&
Think about the two triangles in the figure. What can you say about them? \\

\midrule

\textbf{Focus: Calculate}
&
Ask the student to calculate a value, if necessary, using a calculator.
&
The student has set up an equation or proposed some arithmetic, but did not solve it. Also used to calculate complex values such as $\sin$ and $\arcsin$, which may need a calculator.
&
{\color[HTML]{B7B7B7}That would mean $\sin(x)=0.75$ and $x=\arcsin(0.75)$.}
Okay, can you calculate $x$? You may use a calculator if you want. \\

\midrule

\textbf{Focus: PullBack}
&
Propose that you start solving the question afresh.
&
The student has gone down a very convoluted path that is unlikely to lead to the correct solution.
&
Let's start over and look at the problem from a different angle. What do you notice about the right angles at M and P? \\

\midrule

\textbf{Probing: SelfCorrect}
&
Ask the student to revisit what they just said and try to fix their error.
&
Use when the student's last utterance has an obvious error.
&
Look again at the angles you're adding. Are you sure $\angle$LNM and $\angle$MNP should sum to $90^\circ$? Consider the right angle at M in the context of the whole triangle. \\

\midrule

\textbf{Probing: AskForExplan.}
&
Ask the student to explain their reasoning in more detail.
&
Similar to SelfCorrect, but used when you are not sure what the error is, or whether there is an error.
&
Why did you assume that the sum of $\angle$LNM and $\angle$MNP is $90^\circ$? \\

\midrule

\textbf{Probing: SeekKnowledge}
&
Ask the student to remember some knowledge relevant to the problem but not provided in the problem statement.
&
Use when external knowledge is needed, such as formulae, unit conversions, or simple common-sense knowledge about the world. We recommend writing your own utterance here, as LLMs might be unreliable.
&
Do you remember how much the interior angles of a triangle add up to? \\

\midrule

\textbf{Telling: ExplainConcept}
&
Reveal the method to the answer, but leave the final calculation to the student.
&
Use when everything else fails. Avoid if possible. We recommend writing your own utterance here, as LLMs might be unreliable.
&
So you can see that $\angle$NLP and $\angle$LNP add up to $90^\circ$. Since you already know $\angle$NLP, you can use it to calculate $\angle$LNP. Can you try that? \\

\midrule

\textbf{Telling: Revealing}
&
Reveal the final answer.
&
Use when everything else fails, including ExplainConcept. Avoid if possible. We recommend writing your own utterance here, as LLMs might be unreliable.
&
So you can see that $\angle$NLP and $\angle$LNP add up to $90^\circ$. Since $\angle$NLP is $25^\circ$, we get $\angle$LNP is $65^\circ$. Does that make sense? \\

\bottomrule
\end{tabular}
\end{table*}
}
\section{A Combined Highlight Score}
\label{sec:combined}
Let $H_i$ be the ground truth set of highlights for the $i^{th}$ data point. A data point here refers to a single teacher's utterance to which highlights may be attached. This set can be empty if no highlights are made of the current category for the $i^{th}$ data point.  
Let $\pi_i$ be the output of the corresponding \Pd\ model, which is $1$ if the model wants to highlight or $0$ if it does not. Finally, let $\Delta_i$ be the output of the \D\ model for the $i^{th}$ data point, which is again a set, but cannot be the empty set as per model instructions. 

We start by defining $D_i$, the effective set of highlights, which is identical to $\Delta_i$ if the \Pd\ model outputs a `yes' but the empty set otherwise:
\begin{align*}
    D_i=\begin{cases}
        \Delta_i & if \pi_i=1\\
        \phi & \text{otherwise}
    \end{cases}
\end{align*} 

For the individual data point, the f1 (note the use of lower case f) as:
\[f1_i=\frac{2*|H_i\cap D_i|}{|H_i\cup D_i|+|H_i\cap D_i|}\]
 Normally, this definition is only valid if $|H_i\cap D_i|$ (i.e. true positives) is non-zero, but here we extend it to all cases where $|H_i\cup D_i|$ is defined and non-zero. If however $H_i$ is empty, $\Delta_i$ is undefined, and thus $D_i$ is also undefined. 
In such a case, if $\pi_i$ is $1$, we are saying that the \Pd\ model decided to make some highlight. If, hypothetically, we did run the \D\ model on this example, we would still get some output $\Delta'_i$ with non-zero cardinality. Regardless of the actual members of $\Delta'_i$, $|H_i\cup D'_i|$ would be non-zero and $|H_i\cap D'_i|$ would be 0, giving an $f1=0$. This leaves out the case where $|H_i|=\pi_i=0$. There is no realistic way to define it, but we still set the score to $0$ to simplify later formulations. So we update our definition to be:
\begin{align*}
    f1_i=\begin{cases}
        \frac{2*|H_i\cap D_i|}{|H_i\cup D_i|+|H_i\cap D_i|} & if |H_i|>0\\
        0 & \text{otherwise}
    \end{cases}
\end{align*}

Armed with this definition, we can now proceed to aggregating over data points

We already have the \D\ Accuracy we defined in \Cref{sec:5.1} which aggregates over only those instances where $|H_i|>0$. 
This is like an overall recall, since we are only aggregating over cases where highlights were needed. Let us call this $R$: 
\[
R=\frac{\sum_i f1_i}{\sum_i \mathbb{I}(|H_i|>0)}
\]

A corresponding precision score would act over all cases where $\pi_i=1$. We call this $P$:
\[
P=\frac{\sum_i f1_i}{\sum_i \pi_i}
\]
Note that cases where $|H_i|=\pi_i=0$ never contribute to the denominator, so setting $f1_i=0$ for these cases effectively excludes them from the calculation, as true negatives usually are. Finally, we can define the Combined Highlight Score $F1$ (note the upper case F) as
\[
F1=\frac{2}{\frac{1}{R}+\frac{1}{P}}
\]

\section{Dataset Statistics}
\begin{table*}[t]
\centering
\scriptsize
\setlength{\tabcolsep}{4.5pt}
\renewcommand{\arraystretch}{1.12}
\caption{Comparison of GeoDial with representative educational dialogue datasets. Existing datasets provide valuable text-based tutoring, classroom, or synthetic educational dialogues, but do not capture how teachers coordinate instructional language with explicit visual actions over geometry diagrams.}
\begin{tabular}{p{2.1cm}p{1.9cm}p{3.0cm}p{3cm}p{2.3cm}}
\toprule
\textbf{Dataset} &
\textbf{\#Entries} &
\textbf{Domain} &
\textbf{Interaction Setting} &
\textbf{Modality} \\
\midrule

MathDial~\citep{macina2023mathdial} &
3K dialogs &
Math reasoning problems &
Human teachers tutor LLM-simulated students in one-to-one dialogs. &
Text \\

Book2Dial~\citep{wang-etal-2024-book2dial} &
1,778 dialogs &
Textbook-based education across math, business, science, and social science &
Synthetic teacher--student conversational QA generated from textbook sections. &
Text \\

NCTE~\citep{demszky-hill-2023-ncte} &
1,660 classroom transcripts &
Elementary mathematics &
Classroom instruction from 4th--5th grade math lessons. &
Text \\

CIMA~\citep{stasaski-etal-2020-cima} &
350 shape exercises + 391 prepositional phrase exercises &
Language learning &
Crowdworkers role-play both students and tutors in asynchronous one-to-one tutoring interactions. &
Text + image stimuli \\

\midrule
\textbf{GeoDial (Ours)} &
\textbf{1,341 dialogs} &
\textbf{Geometry problems} &
\textbf{Human math teachers tutor VLM-simulated students in one-to-one dialogs} &
\textbf{Text + geometry diagrams + visual highlights} \\
\bottomrule
\end{tabular}

\label{tab:dataset_comparison}
\end{table*}

\begin{figure}
    \centering
    \includegraphics[width=0.5\linewidth]{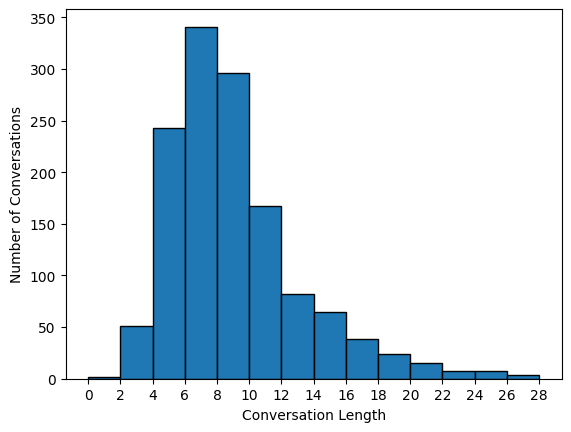}
    \caption{Histogram showing the distribution of conversation lengths in \GD. Only teacher utterances are counted.}
    \label{fig:conv_length}
\end{figure}

\begin{figure}
    \centering
    \includegraphics[width=0.5\linewidth]{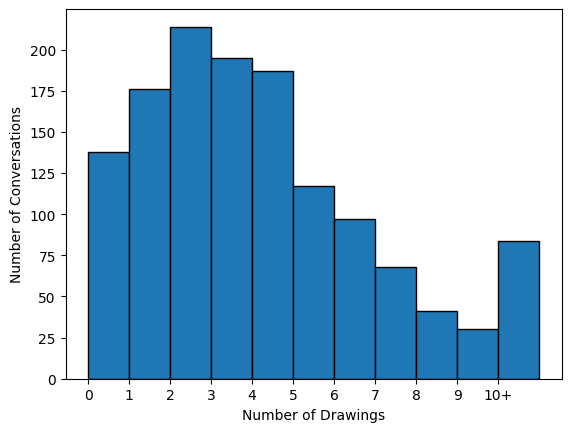}
    \caption{Histogram showing distribution of number of highlighted diagrams per conversation. All conversations with more than 10 diagrams are binned together.}
    \label{fig:images_per_conv}
\end{figure}

\begin{figure}
    \centering
    \includegraphics[width=0.7\linewidth]{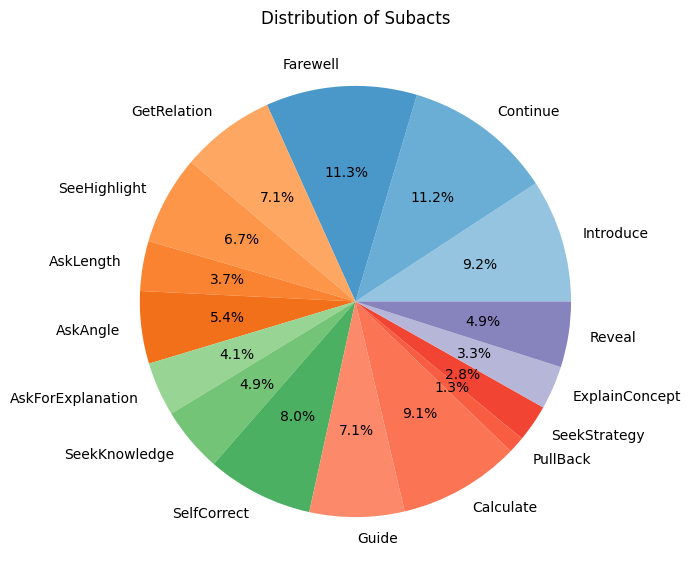}
    \caption{Percentage of each subact in \GD. Subacts belonging to the same act use similar shades}
    \label{fig:subact_per}
\end{figure}

\begin{figure}
    \centering
    \includegraphics[width=0.5\linewidth]{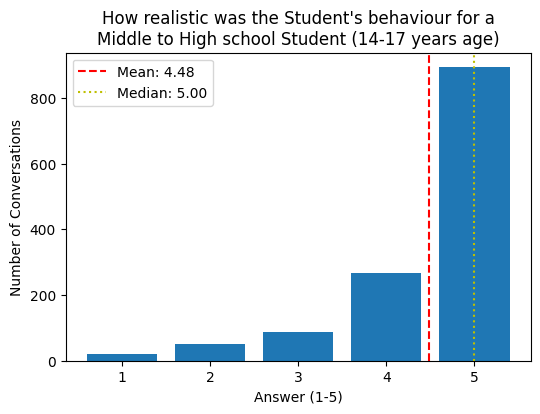}
    \includegraphics[width=0.5\linewidth]{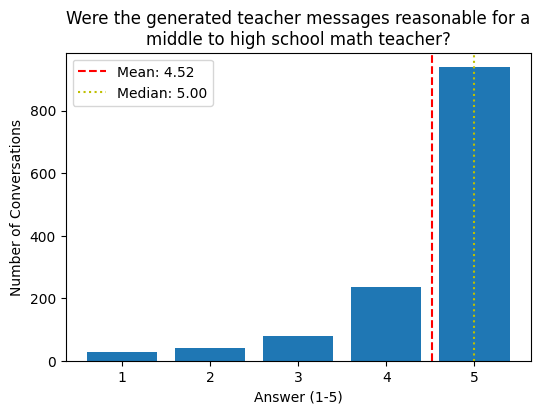}
    \includegraphics[width=0.5\linewidth]{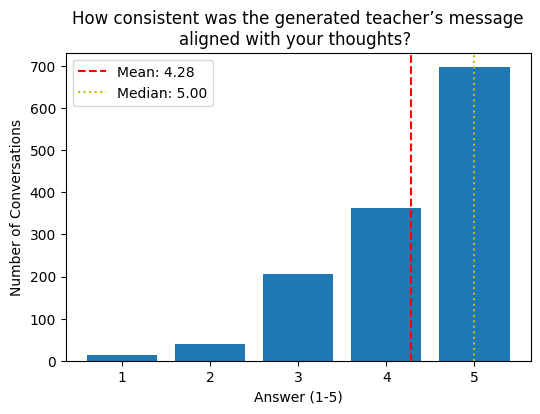}
    \includegraphics[width=0.5\linewidth]{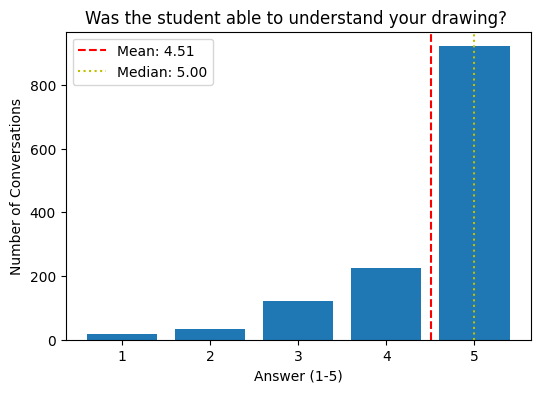}
    \caption{Full distribution of answers to the debrief questions}
    \label{fig:debrief}
\end{figure}

\Cref{fig:conv_length,fig:images_per_conv,fig:subact_per,fig:debrief} show some additional dataset statistics skipped in the main paper due to space constraints. The distribution of both teacher utterances and teacher utterances with highlights follow a chi-squared distribution. Responses to debrief questions show massive peaks at 5, which are always their modes and medians. \Cref{tab:dataset_comparison} compares our \GD\ dataset with representative existing educational dialogue datasets.

\section{Detailed Related Work} \label{sec:related_work}

\subsection{Educational Dialog Datasets}
The development of educational tutoring systems has been driven by the availability of educational dialog datasets. Recent efforts have focused on collecting text-only tutoring conversations, such as MathDial~ \citep{macina2023mathdial}, where teachers interacted with LLM-simulated students to produce pedagogically grounded instructional dialogs.
Several other datasets were created by having crowdworkers role-play both teacher and student, such as CIMA~ \citep{stasaski-etal-2020-cima}, by transcribing classroom interactions or logging online conversations~ \citep{suresh2022fine, demszky-hill-2023-ncte,caines2020teacher}, or by using models to simulate teacher and student \citep{wang-etal-2024-book2dial}.
However, these resources are primarily text-based and do not capture teacher visual actions over diagrams, despite the heavy use of diagrams in math education. In contrast, \GD\ provides dialogs that integrate diagram interpretation and fine-grained diagram highlight annotations by teachers.

\subsection{Multimodal Geometry Datasets}


Research on multimodal geometry datasets has primarily aimed to enable models to solve geometry problems by integrating textual descriptions with diagrammatic information. 
Influential work by Seo et al.~ \citep{Seo_Hajishirzi_Farhadi_Etzioni_2014,seo-etal-2015-solving} established diagram understanding as a core challenge in geometry problem solving, introducing structured representations that jointly model visual elements, spatial relations, and their alignment with textual descriptions for geometry problem solving. 
This led to several larger datasets and efforts of symbolic and later neural reasoners for geometry problems. 
For instance, Geometry3K~ \citep{lu2021inter} introduced the notion of diagram components and solution programs to support rule-based geometry reasoning. 
Subsequent works further increased annotation granularity: PGPS9K \citep{PGPS9k} introduced fine-grained parsing and structural labels to support reasoning over geometric elements.
Another direction of research in geometry problem solving focused on coordinate geometry driven numerical calculation. 
For instance, GeoQA~ \citep{chen-etal-2021-geoqa} demonstrated that pairing diagrams with question–answer supervision enables numerical reasoning over visual geometry problems. 
UniGeo~ \citep{chen-etal-2022-unigeo} spanned both calculation and proving, including tasks such as computing missing angles and proving triangle congruence.
Recent work has also focused on scaling multimodal geometry resources for model training. 
Geo170K~ \citep{ICLR2025_09afabe3}, for example, provides a large corpus of automatically generated diagram-instruction pairs, while emerging datasets such as GeoSense~ \citep{xu2025geosense} explore conceptual reasoning by evaluating whether models can recognize and apply geometric principles.
Despite these advances, existing multimodal geometry datasets do not capture how teachers interact with diagrams to guide student learning. \GD\  addresses this gap by offering multimodal tutoring dialogs grounded in geometry diagrams, with explicit annotations that connect visual cues to instructional intent.

\subsection{Visuals in Mathematics Education}
Visuals have long been recognized as critical tools in mathematics education~ \citep{kaitera2022developing,boonen2016s}. Research indicates that visual representations help students grasp abstract concepts more readily~ \citep{small2025eyes,MAYER200285,evagorou2015role}, increase engagement~ \citep{cooper2018benefits}, and improve learning efficiency~ \citep{arcavi2003role}.
Within geometry education in particular, visual cues such as highlighted segments and annotated angles have been shown to support students in learning geometry concepts~ \citep{SCHOENHERR2024100639} and help them grasp geometric relationships~ \citep{jones2013diagrams}.

Despite this pedagogical importance, existing educational datasets and model training pipelines rarely incorporate visual grounding. \GD\ embeds visual feedback directly into the tutoring interaction: teacher utterances reference highlighted regions of diagrams, and the dataset provides structured visual annotations that enable models to learn to generate diagram highlights.
\section{Filtering the Dialogs}
\label{sec:filter}
We filtered out some dialogs which had quality issues.
Most of the errors in the data collection were due to network issues, causing errors with the order in which utterances were shown to the annotator and the order in which they were presented to the student LLM. While such dialogs usually still make sense, it is unclear what the correct order would be, and hence we decided to drop any conversations where this happened. These errors would manifest as either the interleaving of two conversations, or consecutive student and/or teacher turns. The interleaving happened in 25 of the 237 batches we collected, and these files were manually inspected by the first author to keep only clean conversations. A further 25 conversations were dropped due to consecutive turns. Finally, for one batch, the highlighted diagrams did not get saved on our server due to reasons unclear to us. This was also dropped. 
\section{Annotator Interviews for Assessing Faithfulness of Student Simulation}
\label{sec:inter}
Our interview, inspired by \citet{martynova-etal-2025-llms}, consisted of 6 qualitative questions, each followed by a rating on a 5-point Likert scale. These questions, in order, dealt with student engagement, student emotions, diagram/highlight understanding, error quality, teacher frustration and teacher strategy (the full text is provided in \Cref{tab:questions}). Interviewees were asked to compare their experience in teaching and tutoring real students with their experience in \GD. For the first 3 questions, the Likert rating went from 1=\GD\ students show better engagement/ more emotions/ better diagram understanding, to 5=Real students show better engagement/ more emotions/ better diagram understanding. For teacher frustration the scale is reversed since it usually hurts teaching \citep{buric2023teacher}. Error quality discussed  how realistic the student wrong solutions were, and the scale went from 1=Indistinguishable from real student errors to 5=Very unrealistic. Finally, the teaching strategy question tried to understand if the general strategy the interviewees used for \GD\ was any different from what they would use with actual students and the Likert rating went from 1=No change in strategy to 5=Very different strategy. The interviews ended with a general feedback question asking if the interviewees wanted to suggest any changes in the setup. The full set of questions is provided in \Cref{tab:questions}.

All interviewees were initially asked to talk about their teaching experiences. Four of them were regular teachers having 3, 5, 6 and  10 years of teaching experience, having taught approximately 150, 450, 500, and 300 unique students over their career, respectively. The fifth interviewee has worked with a homework support website, where students can upload their homework questions to get paired with tutors who can help them online. He has been working for about 9 years and has had over 30,000 individual sessions. Although this is not the same as classroom teaching, we believe it is rather close to our setup.

The interviewees believed engagement to be slightly better for the AI students (Ratings= 2/2/2/2/3) used in \GD, which is not surprising since AI models, unlike real students, cannot opt to disengage or leave the conversation. Our definition of engagement included contesting the teachers' claims and explanations, which multiple interviewees felt was quite common in the AI. On emotions, however, AI students lose out (Ratings=3/3/4/5/5). We should note here that at least 2 interviewees preferred emotionless students as they felt that the lack of student emotions would make them better teachers. The student confusions were seen as very realistic (Ratings=1/1/1/2/2), though 2 interviewees noted that on some rare occasions, the dialog would sometimes show inconsistencies with the original confusion. All 3 of these findings are consistent with that of \citet{martynova-etal-2025-llms}. Teacher frustration ratings (1/2/3/4/5) were all over the board, with interviewees being split on whether or not frustration makes them perform better.  

Diagram understanding is the primary innovation of our work, and it is important that simulated students demonstrate understanding of teachers' visual cues.
While the ratings (1/3/3/4/4) say that the AI students matched real students, the discussions in the interviews seemed to lean very slightly towards real students. 
The vastly different ways diagrams were used by the different interviewees meant that the human reference kept changing.
The interviewees were split on whether or not giving the students the ability to highlight is important, mostly based on whether or not their way of teaching allows for students to interact with the diagram.  While this is a limitation we did not address, the interviews show that one-way diagram interaction does happen in the real world. 

Finally, the interviewees believed that they only slightly changed their teaching strategy (Ratings=1/2/2/2/3) from their interactions with real students. The main driver behind this seemed to be the uniqueness of the setting, as even in one-to-one tutoring scenarios, solution errors are interspersed with concept revision, and correctly solved questions. The worst rating of 3 came from the interviewee who claimed to use questions only as a diagnostic tool to identify concept gaps, and then addressing those gaps instead of getting the student to fix their solution. As we stated earlier, the entire setup of \GD\ is in service of extracting realistic tutoring strategies from teachers, making this the most important factor behind the fidelity of \GD. While it is not perfect, we believe that a mean rating of 2 (Slight change in strategy) shows that we were quite successful in our goal.
\section{Translating Text Annotations into Diagram Highlights}
\label{sec:6.2}

Our proposed \GD\ dataset supports multiple modeling paradigms for building multimodal tutoring systems. The most direct approach is to train a unified model that both understands the diagram and generates a new image containing the highlight. However, this type of full-image generation is typically slow, computationally expensive, and prone to visual hallucinations~ \citep{aithal2024,ma2025scaling,wang-etal-2025-generating-pedagogically}. 
These issues make full image generation unsuitable for real-time tutoring or lightweight deployment.
Thus, as we mentioned in \Cref{sec:ann}, we opt to visual highlights to textual highlight annotations, which we use all across \Cref{sec:7}. However, in order to deploy in the real world, we would have to recover the visual highlights from generated textual highlights. 
In this section, we introduce an automatic drawing pipeline that achieves this.

\subsection{Vertex and Label Detection}\label{sec:NLD}
Vertex detection identifies the pixel location of every geometric point referenced in a highlight instruction. The system first parses the diagram using GeoSolver~ \citep{seo-etal-2015-solving}, which extracts structural point candidates from intersections and curve endpoints across all sub-diagram regions. 
We then render a temporary version of the diagram that overlays each detected point with a unique numeric marker (see example in~\Cref{fig:numeric_overlay}). This is fed to Gemini, which identifies the correspondence between each printed letter in the original diagram and the numeric overlay, establishing a mapping from letter labels to precise pixel coordinates.

For vertices that cannot be resolved through the matching procedure, DeepSeek-OCR~ \citep{wei2025deepseek} is applied to directly detect printed letter labels on the diagram and provide their bounding box coordinates. DeepSeek-OCR is also used to handle textual references that are not tied to vertices, e.g., side lengths, variable names, etc., which are often marked in the ``Labels'' field mentioned above.


\subsection{Highlights Drawing}
Once we have the coordinates and bounding boxes, we can use them to highlight the figure. The renderer supports segment traces for lines, ray-based angle markings, circle-constrained arcs, and circular emphasis for textual labels, as shown in \Cref{sec:example_highlight}. 
A set of geometric adjustment procedures manages visual crowding: collinear lines are shifted to avoid overlap, and angles sharing a vertex are displaced along bisectors. These steps produce clear, interpretable highlights, ensuring that the rendered annotations remain legible even under complex instructions.

\subsection{Evaluation} \label{eval_detail}
We evaluated the automatic drawing pipeline through a human evaluation on 100 randomly sampled utterances from the \GD\ dataset. Two researchers independently reviewed the rendered diagrams and then discussed any disagreements to reach a consensus. Each sample was scored on a three-level scale: a score of 1.0 was assigned when all rendered highlights aligned with the instruction; 0.5 was assigned when the instruction contained multiple required highlights, and only a subset was rendered correctly; and 0.0 was assigned when none of the required highlights were rendered correctly. Under this evaluation protocol, the pipeline achieved an accuracy of 95\%, demonstrating that symbolic rendering effectively translates textual highlight descriptions into precise diagram annotations.
\section{Model Details}\label{app:e}
In the main body of the paper, we often referred to LLMs and VLMs by their generic name to avoid complicating the narrative. In this section, we present full details of the models used. 

The most commonly used model is \texttt{gemini-2.5-flash-preview-05-20} \citep{googleGeminiFlash}, which we refer to as ``Gemini''. This model is one of the Large models used for suggesting utterances, and was used as a test case in  \Cref{tab:results}. It is also used to filter out diagrams lacking point names and  generating ground truth solutions in  \Cref{sec:PSE} and for detection of node correspondence in  \Cref{sec:NLD}. This also acts as the student model in the annotation setup. A different version of Gemini, \texttt{Gemini-1.5-Flash-8B} was used as the smaller model for generating student errors in  \Cref{sec:PSE}, though this was deprecated while we were halfway through our data collections, and we had to switch to a different model.

The other two large models were \texttt{claude-sonnet-4-20250514} \citep{anthropic2025claude4systemcard} usually referred to as ``Claude'' and \texttt{gpt-4o} \citep{gpt} usually referred to as ``GPT'' in the text. Both models were used to suggest teacher utterances and for evaluation in  \Cref{tab:results}.

We sourced all our open models from HuggingFace\footnote{\href{https://huggingface.co/}{https://huggingface.co/}} and shall provide model names as they appear on HuggingFace. Qwen3  \citep{bai2025qwen3vltechnicalreport} was the most used model in this paper. When \texttt{Gemini-1.5-Flash-8B} was deprecated, we switched to using \texttt{Qwen/Qwen3-VL-8B-Instruct} to generate student errors in  \Cref{sec:PSE}. This model was also used for evaluation in  \Cref{tab:results}. This table also uses 3 other versions of Qwen3, namely the text-only models \texttt{Qwen/Qwen3-32B} and \texttt{Qwen/Qwen3-8B} and the VLM \texttt{Qwen/Qwen3-VL-32B-Instruct}. Other than this, we used two Llama models \citep{grattafiori2024llama3herdmodels}, namely \texttt{meta-llama/Llama-3.1-8B-Instruct} and \texttt{meta-llama/Llama-3.2-11B-Vision}, and one InternVL \citep{wang2025internvl3} model.
\texttt{OpenGVLab/InternVL3\_5-8B-Instruct}.  All models in \Cref{tab:results} are finetuned over the train set for $5$ epochs, with a batch size of 8,  and learning rate 1e-4 selected based on prior studies and pilot experiments that showed stable convergence. To ensure parameter efficiency, we incorporate LoRA adapters~ \citep{hu2022lora}, which significantly reduce the number of trainable parameters while maintaining performance. All models were fine-tuned on four NVIDIA RTX 4090 GPUs, and required 32 hours of training.

We fine-tuned all models, including Qwen3 (8B and 32B), Qwen3-VL (8B and 32B Instruct), LLaMA-3.1 (8B), LLaMA-3.2 (11B Vision), and InternVL-3.5 (8B Instruct), under a unified training setup using the ms-swift~\citep{zhao2025swift} framework. Training was conducted on four NVIDIA GH200 GPUs with LoRA-based supervised fine-tuning. We used \texttt{bfloat16} precision, flash attention, padding-free packed sequences, and gradient checkpointing to reduce memory usage while supporting long contexts up to 8,192 tokens. All models were trained for five epochs with a per-device batch size of one and gradient accumulation over two steps. Optimization used AdamW with a learning rate of $1\times10^{-4}$ and a warm-up ratio of 0.05. LoRA adapters were applied to all linear layers with rank 8 and scaling factor 32, while the vision encoder and aligner modules were frozen. Each training run took approximately 6 to 15 hours. This unified configuration ensures fair comparison across models while maintaining training efficiency and stability.

\section{Additional Experiment Results Tables}


\begin{table}[t]
\centering
\footnotesize
\setlength{\tabcolsep}{4.5pt}
\renewcommand{\arraystretch}{1.08}
\caption{Change in prediction precision and recall for VLMs. Values in parentheses show the change from the corresponding model without finetuning.}
\label{tab:prc_rec}

\newcolumntype{C}[1]{>{\centering\arraybackslash}p{#1}}

\begin{tabular}{
C{2.3cm}
C{1.4cm}
@{\hspace{5pt}}|@{\hspace{5pt}}
ccc
@{\hspace{5pt}}|@{\hspace{5pt}}
ccc
}
\toprule
\multirow{2}{*}{\textbf{Model}} &
\multirow{2}{*}{\textbf{Finetuning}} &
\multicolumn{3}{c@{\hspace{5pt}}|@{\hspace{5pt}}}{\textbf{Line Prediction (\%)}} &
\multicolumn{3}{c}{\textbf{Angle Prediction (\%)}} \\
\cmidrule(lr){3-5}
\cmidrule(lr){6-8}
& &
\textbf{Prec.} & \textbf{Rec.} & \textbf{F1} &
\textbf{Prec.} & \textbf{Rec.} & \textbf{F1} \\
\midrule

\multirow{2}{*}{Qwen-3-VL-8B}
& Without
& 87.0 & 29.3 & 43.8
& 27.8 & 97.6 & 43.3 \\
& With
& \makecell{42.5\\{\scriptsize (-44.5)}}
& \makecell{62.1\\{\scriptsize (+32.8)}}
& \makecell{50.5\\{\scriptsize (+7.7)}}
& \makecell{45.6\\{\scriptsize (+17.8)}}
& \makecell{33.2\\{\scriptsize (-64.4)}}
& \makecell{38.4\\{\scriptsize (-4.9)}} \\

\midrule

\multirow{2}{*}{Llama-3.2-11B-V}
& Without
& 84.1 & 28.7 & 42.8
& 22.8 & 38.3 & 28.6 \\
& With
& \makecell{39.0\\{\scriptsize (-45.1)}}
& \makecell{65.2\\{\scriptsize (+36.5)}}
& \makecell{48.8\\{\scriptsize (+6.0)}}
& \makecell{45.7\\{\scriptsize (+22.9)}}
& \makecell{35.9\\{\scriptsize (-2.4)}}
& \makecell{40.2\\{\scriptsize (+11.6)}} \\

\midrule

\multirow{2}{*}{Qwen-3-VL-32B}
& Without
& 87.0 & 29.6 & 44.2
& 23.1 & 99.5 & 37.5 \\
& With
& \makecell{41.6\\{\scriptsize (-45.4)}}
& \makecell{64.6\\{\scriptsize (+35.0)}}
& \makecell{50.6\\{\scriptsize (+6.4)}}
& \makecell{50.5\\{\scriptsize (+27.4)}}
& \makecell{41.8\\{\scriptsize (-57.7)}}
& \makecell{45.8\\{\scriptsize (+8.7)}} \\

\bottomrule
\end{tabular}
\end{table}
\begin{table}[t]
\centering
\footnotesize
\setlength{\tabcolsep}{5.0pt}
\renewcommand{\arraystretch}{1.05}
\caption{Exploring the cause behind a 0 Decision F1 score. We split these cases into misses, where there are false positives but no true positives, and leaves, where there are neither false positives nor true positives. Finetuning reduces the miss rate but increases the leave rate.}
\label{tab:missleave}

\newcolumntype{C}[1]{>{\centering\arraybackslash}p{#1}}

\begin{tabular}{
C{2.5cm}
C{1.6cm}
@{\hspace{5pt}}|@{\hspace{5pt}}
cc
@{\hspace{5pt}}|@{\hspace{5pt}}
cc
}
\toprule
\multirow{2}{*}{\textbf{Model}} &
\multirow{2}{*}{\textbf{Finetuning}} &
\multicolumn{2}{c@{\hspace{5pt}}|@{\hspace{5pt}}}{\textbf{Line (\%)}} &
\multicolumn{2}{c}{\textbf{Angle (\%)}} \\
\cmidrule(lr){3-4}
\cmidrule(lr){5-6}
& &
\textbf{Miss} & \textbf{Leave} &
\textbf{Miss} & \textbf{Leave} \\
\midrule

\multirow{2}{*}{Qwen-3-VL-8B}
& Without & 38.3 & 2.3  & 64.5 & 1.8  \\
& With    & 10.4 & 57.5 & 17.7 & 66.8 \\

\midrule

\multirow{2}{*}{Llama-3.2-11B-V}
& Without & 39.0 & 11.4 & 44.5 & 46.8 \\
& With    & 10.4 & 61.0 & 19.1 & 64.1 \\

\midrule

\multirow{2}{*}{Qwen-3-VL-32B}
& Without & 34.4 & 0.3  & 50.5 & 0.5  \\
& With    & 10.4 & 58.4 & 25.0 & 58.2 \\

\bottomrule
\end{tabular}
\end{table}
\begin{table}[t]
\centering
\footnotesize
\setlength{\tabcolsep}{5.0pt}
\renewcommand{\arraystretch}{1.05}
\caption{Ablation on training set size.}
\label{tab:data_size}

\newcolumntype{C}[1]{>{\centering\arraybackslash}p{#1}}

\begin{tabular}{
C{1.7cm}
@{\hspace{5pt}}|@{\hspace{5pt}}
cccc
}
\toprule
\textbf{\# Dialogs} &
\multicolumn{4}{c}{\textbf{Performance}} \\
\cmidrule(lr){2-5}
\textbf{Trained On} &
\textbf{Act} &
\textbf{Subact} &
\textbf{Feedback} &
\textbf{Utterance} \\
&
\textbf{Macro-F1} &
\textbf{Macro-F1} &
\textbf{Macro-F1} &
\textbf{BLEU} \\
\midrule
0         & 32.2 & 13.3 & 54.3 & 3.2 \\
200       & 41.0 & 24.4 & 68.1 & 11.2 \\
400       & 43.7 & 28.3 & 69.4 & 11.7 \\
600       & \textbf{45.4} & \textbf{30.2} & 69.5 & 11.6 \\
800 (Max) & 45.3 & \textbf{30.2} & \textbf{70.0} & \textbf{12.4} \\
\bottomrule
\end{tabular}
\end{table}

\begin{table*}[t]
\centering
\setlength{\tabcolsep}{3.0pt}
\caption{95\% Confidence intervals of focused experiments on highlighting. ``From Model'' indicates that provided teacher utterances come from a Qwen-VL-8b model trained to generate teacher utterances without seeing the ground truth highlights. Combined Scores are calculated in accordance with \Cref{sec:combined}.}
\label{tab:focus_c1}
\resizebox{\textwidth}{!}{
\begin{tabular}{ccc|cccc|ccc|ccc}
\toprule
\textbf{Element} &
\textbf{Teacher Utterance} &
\textbf{Model} &
\multicolumn{4}{c|}{\textbf{Prediction (\%)}} &
\multicolumn{3}{c|}{\textbf{Decision, Independent (\%)}} &
\multicolumn{3}{c}{\textbf{Combined Score (\%)}} \\
\cmidrule(lr){4-7}
\cmidrule(lr){8-10}
\cmidrule(lr){11-13}
& & &
\textbf{Prec.} & \textbf{Rec.} & \textbf{F1} & \textbf{Acc.} &
\textbf{Prec.} & \textbf{Rec.} & \textbf{F1} &
\textbf{Prec.} & \textbf{Rec.} & \textbf{F1} \\
\midrule

\multirow{7}{*}{\textbf{Line}}
& \multirow{3}{*}{From Model}
& Weighted
& (72.8,80.0)& (54.9,62.3)& (62.9,69.0)& (83.9,86.7)& \multicolumn{3}{c|}{Same as fine-tuned model}
& (45.4,53.2)& (34.3,40.8)& \textbf{(39.2,45.7)}\\
& & Yes
& (65.9,73.6)& (53.6,61.1)& (59.4,65.8)& (81.9,84.7)& (61.4,68.0)& (60.6,67.0)& (58.4,64.5)& (41.8,49.4)& (34.1,40.8)& (37.7,44.2)\\
& & No
& (37.6,42.5)& (84.1,89.2)& (52.1,57.1)& (62.4,66.0)& (59.4,66.0)& (56.7,63.1)& (55.6,61.8)& (21.2,25.1)& (47.1,53.6)& (29.3,33.9)\\
\cmidrule(lr){2-13}

& \multirow{2}{*}{Ground Truth}
& Yes
& (76.0,82.6)& (63.8,70.8)& (69.7,75.2)& (86.3,88.7)& (76.7,82.2)& (77.4,82.9)& (75.4,80.8)& (64.5,71.5)& (54.2,61.1)& (59.1,65.2)\\
& & No
& (37.7,42.6)& (83.3,88.7)& (52.0,57.0)& (52.0,57.0)& (76.8,82.3)& (72.4,78.0)& (72.7,78.1)& (28.5,32.9)& (62.7,69.1)& (39.2,44.2)\\
\cmidrule(lr){2-13}

& \multirow{2}{*}{Not Given}
& Yes
& (67.7,75.3)& (53.2,60.7)& (59.9,66.1)& (82.3,85.1)& (61.3,67.7)& (63.2,69.7)& (59.7,65.9)& (42.5,50.0)& (33.5,40.0)& (37.5,44.0)\\
& & No
& (38.6,43.8)& (81.7,87.3)& (52.6,57.8)& (64.2,67.7)& (49.7,56.7)& (42.4,48.7)& (42.8,48.8)& (16.4,19.9)& (34.4,40.6)& (22.2,26.5)\\

\midrule

\multirow{7}{*}{\textbf{Angle}}
& \multirow{3}{*}{From Model}
& Weighted
& (59.1,68.5)& (49.6,58.5)& (54.3,61.9)& (85.2,87.8)& \multicolumn{3}{c|}{Same as fine-tuned model}
& (27.5,35.8)& (23.1,30.4)& \textbf{(25.2,32.4)}\\
& & Yes
& (57.4,66.4)& (53.7,62.5)& (55.8,63.2)& (85.0,87.5)& (46.1,54.3)& (45.5,53.5)& (44.9,52.7)& (25.6,33.2)& (23.9,31.2)& (24.8,31.8)\\
& & No
& (31.2,36.4)& (80.0,86.9)& (45.0,50.8)& (66.1,69.6)& (41.7,49.8)& (42.0,50.3)& (40.9,48.8)& (12.7,16.3)& (32.3,40.0)& (18.3,23.0)\\
\cmidrule(lr){2-13}

& \multirow{2}{*}{Ground Truth}
& Yes
&  (72.2,80.7)&  (60.1,68.6)& (66.0,72.9)& (89.1,91.3)& (59.6,67.3)& (59.2,67.0)& (58.7,66.3)& (46.1,55.1)& (38.3,46.5)& (42.0,49.9)\\
& & No
& (34.7,40.2)& (86.2,91.9)& (49.7,55.3)& (69.6,73.0)& (55.6,63.9)& (54.7,63.0)& (54.4,62.5)& (19.5,24.0)& (48.0,56.2)& (27.7,33.4)\\
\cmidrule(lr){2-13}

& \multirow{2}{*}{Not Given}
& Yes
& (59.0,67.9)& (54.2,62.9)& (56.8,64.1)& (85.5,87.9)& (39.5,47.8)& (40.0,48.3)& (38.7,46.7)& (24.4,32.0)& (22.3,29.4& (23.4,30.4)\\
& & No
& (32.5,37.9)& (77.7,84.8)& (46.0,51.9)& (68.2,71.6)& (42.0,50.2)& (40.2,48.3)& (39.6,47.3)& (10.8,16.2)& (29.7,37.2)& (17.6,22.4)\\

\bottomrule
\end{tabular}
}
\end{table*}
\begin{table*}[t]
\centering
\scriptsize
\setlength{\tabcolsep}{2.3pt}
\renewcommand{\arraystretch}{0.92}
\caption{95\% Confidence Intervals of models on \GD\ validation set, for Categorical variables, Decision F1 and Utterance quality. Models that do not have access to question image are in \textit{{\colorbox[HTML]{E4DFF4}{lavender italics}}}. Confidence Intervals are obtained by bootstrapping the sample with a sample size of $10000$.}
\label{tab:results_val_ci1}
\resizebox{\textwidth}{!}{
\begin{tabular}{clccc|cccc|cc}
\toprule
\textbf{Group} &
\textbf{Model} &
\multicolumn{3}{c|}{\textbf{Categorical Macro-F1}} &
\multicolumn{4}{c|}{\textbf{Decision F1}} &
\multicolumn{2}{c}{\textbf{Utterance}} \\
\cmidrule(lr){3-5}
\cmidrule(lr){6-9}
\cmidrule(lr){10-11}
& &
\textbf{Act} & \textbf{Subact} & \textbf{Feedback}
& \textbf{Line} & \textbf{Angle} & \textbf{Arc} & \textbf{Label}
& \textbf{BERT} & \textbf{BLEU} \\
\midrule

\multirow{1}{*}{\rotatebox[origin=c]{90}{}}
& Random Turn
& (24.7, 29.3) & (13.4, 17.0) & (44.0, 49.6) & (0.4, 1.7) & (0.0, 0.7) & (0.0, 0.0) & (0.0, 2.4) & (12.9, 15.2) & (3.1, 4.3)\\

\midrule

\multirow{3}{*}{\rotatebox[origin=c]{90}{\textbf{Closed}}}
& Gemini 2.5 Flash
& (29.4, 35.3) & (13.8, 17.2) & (56.0, 61.8) & (19.2, 23.9) & (10.4, 15.6) & (0.0, 3.9) & (6.0, 15.5) & (14.3, 16.2) & (3.0, 3.9) \\

& GPT-4o
&(26.6, 32.0) & (12.8, 17.4) & (49.4, 55.2) & (25.1, 29.5) & (8.2, 13.0) & (0.0, 0.0) & (0.0, 2.4) & (19.5, 21.9) & (3.0, 3.9) \\

& Claude Sonnet 4
& (29.9, 35.2) & (17.4, 22.5) & (54.9, 61.0) & (13.2, 18.0) & (4.9, 9.2) & (0.0, 0.0) & (0.8, 6.6) & (24.4, 26.3) & (5.5, 6.7) \\

\midrule

\multirow{7}{*}{\rotatebox[origin=c]{90}{\textbf{Open}}}
& {\cellcolor[HTML]{E4DFF4}\textit{Qwen 3 8B}}
& {\cellcolor[HTML]{E4DFF4}\textit{(24.9, 30.1) }}&{\cellcolor[HTML]{E4DFF4}\textit{ (8.7, 13.4) }}&{\cellcolor[HTML]{E4DFF4}\textit{ (43.1, 48.7) }}&{\cellcolor[HTML]{E4DFF4}\textit{ (20.8, 25.1) }}&{\cellcolor[HTML]{E4DFF4}\textit{ (10.0, 14.9) }}&{\cellcolor[HTML]{E4DFF4}\textit{ (0.0, 0.0) }}&{\cellcolor[HTML]{E4DFF4}\textit{ (0.0, 2.5) }}&{\cellcolor[HTML]{E4DFF4}\textit{ (15.6, 17.2) }}&{\cellcolor[HTML]{E4DFF4}\textit{ (1.5, 2.0)}} \\

& {\cellcolor[HTML]{E4DFF4}\textit{Llama 3.1 8B}}
& {\cellcolor[HTML]{E4DFF4}\textit{(17.3, 23.4) }}&{\cellcolor[HTML]{E4DFF4}\textit{ (5.4, 12.0) }}&{\cellcolor[HTML]{E4DFF4}\textit{ (41.0, 46.7) }}&{\cellcolor[HTML]{E4DFF4}\textit{ (14.0, 18.2) }}&{\cellcolor[HTML]{E4DFF4}\textit{ (2.5, 5.7) }}&{\cellcolor[HTML]{E4DFF4}\textit{ (0.0, 0.0) }}&{\cellcolor[HTML]{E4DFF4}\textit{ (0.0, 2.4) }}&{\cellcolor[HTML]{E4DFF4}\textit{ (17.5, 19.2) }}&{\cellcolor[HTML]{E4DFF4}\textit{ (2.6, 3.4)}} \\

& {\cellcolor[HTML]{E4DFF4}\textit{Qwen 3 32B}}
& {\cellcolor[HTML]{E4DFF4}\textit{(27.9, 32.2) }}&{\cellcolor[HTML]{E4DFF4}\textit{ (10.8, 14.2) }}&{\cellcolor[HTML]{E4DFF4}\textit{ (47.8, 53.9) }}&{\cellcolor[HTML]{E4DFF4}\textit{ (22.3, 26.7) }}&{\cellcolor[HTML]{E4DFF4}\textit{ (11.4, 16.5) }}&{\cellcolor[HTML]{E4DFF4}\textit{ (0.0, 6.1) }}&{\cellcolor[HTML]{E4DFF4}\textit{ (0.4, 3.8) }}&{\cellcolor[HTML]{E4DFF4}\textit{ (18.5, 20.1) }}&{\cellcolor[HTML]{E4DFF4}\textit{ (2.5, 3.2)}} \\

& Qwen 3 VL 8B
& (29.0, 35.4) & (11.0, 15.6) & (51.2, 57.4) & (18.8, 23.1) & (10.2, 15.1) & (0.0, 0.0) & (3.6, 11.5) & (19.0, 20.7) & (2.8, 3.5) \\

& InternVL 8B
(2.8, 6.7) & (0.8, 4.0) & (10.7, 15.5) & (3.6, 6.5) & (1.4, 4.3) & (0.0, 0.0) & (0.0, 0.0) & (2.6, 3.8) & (0.3, 0.7)\\

& Llama 3.2 11B-V
& (15.5, 21.5) & (4.4, 10.1) & (39.2, 44.7) & (14.5, 18.6) & (2.0, 5.0) & (0.0, 0.0) & (0.0, 2.4) & (17.5, 19.1) & (2.4, 3.1) \\

& Qwen 3 VL 32B
& (27.6, 32.9) & (12.0, 16.4) & (53.2, 59.8) & (21.7, 26.1) & (15.4, 20.6) & (0.0, 0.0) & (2.6, 10.1) & (19.1, 20.9) & (3.3, 4.1)\\

\midrule

\multirow{7}{*}{\rotatebox[origin=c]{90}{\textbf{Fine-tuned}}}
& {\cellcolor[HTML]{E4DFF4}\textit{Qwen 3 8B}}
& {\cellcolor[HTML]{E4DFF4}\textit{(41.7, 47.5) }}&{\cellcolor[HTML]{E4DFF4}\textit{ (26.3, 30.7) }}&{\cellcolor[HTML]{E4DFF4}\textit{ (65.4, 71.1) }}&{\cellcolor[HTML]{E4DFF4}\textit{ (10.0, 14.4) }}&{\cellcolor[HTML]{E4DFF4}\textit{ (5.0, 9.2) }}&{\cellcolor[HTML]{E4DFF4}\textit{ (0.0, 0.0) }}&{\cellcolor[HTML]{E4DFF4}\textit{ (1.3, 8.2) }}&{\cellcolor[HTML]{E4DFF4}\textit{ (30.5, 33.1) }}&{\cellcolor[HTML]{E4DFF4}\textit{ (11.3, 13.4)}} \\

& {\cellcolor[HTML]{E4DFF4}\textit{Llama 3.1 8B}}
& {\cellcolor[HTML]{E4DFF4}\textit{(42.3, 48.0) }}&{\cellcolor[HTML]{E4DFF4}\textit{ (27.3, 33.0) }}&{\cellcolor[HTML]{E4DFF4}\textit{ (65.5, 71.1) }}&{\cellcolor[HTML]{E4DFF4}\textit{ (10.2, 14.4) }}&{\cellcolor[HTML]{E4DFF4}\textit{ (3.8, 7.5) }}&{\cellcolor[HTML]{E4DFF4}\textit{ (0.0, 0.0) }}&{\cellcolor[HTML]{E4DFF4}\textit{ (0.0, 4.0) }}&{\cellcolor[HTML]{E4DFF4}\textit{ (29.5, 32.1) }}&{\cellcolor[HTML]{E4DFF4}\textit{ (10.6, 12.5)}} \\

& {\cellcolor[HTML]{E4DFF4}\textit{Qwen 3 32B}}
& {\cellcolor[HTML]{E4DFF4}\textit{(45.2, 51.1) }}&{\cellcolor[HTML]{E4DFF4}\textit{ (28.2, 32.9) }}&{\cellcolor[HTML]{E4DFF4}\textit{ (69.5, 74.9) }}&{\cellcolor[HTML]{E4DFF4}\textit{ (9.5, 13.7) }}&{\cellcolor[HTML]{E4DFF4}\textit{ (4.7, 8.8) }}&{\cellcolor[HTML]{E4DFF4}\textit{ (1.3, 10.5) }}&{\cellcolor[HTML]{E4DFF4}\textit{ (0.0, 3.6) }}&{\cellcolor[HTML]{E4DFF4}\textit{ (31.7, 34.3) }}&{\cellcolor[HTML]{E4DFF4}\textit{ (11.9, 13.9)}} \\

& Qwen 3 VL 8B
& (42.4, 48.3) & (27.5, 33.1) & (66.9, 72.5) & (10.6, 15.0) & (4.4, 8.5) & (0.0, 3.9) & (0.0, 5.6) & (30.5, 33.1) & (11.4, 13.4) \\

& InternVL 8B
& (43.0, 48.9) & (28.9, 34.9) & (67.7, 73.2) & (11.4, 16.0) & (5.3, 9.7) & (0.0, 3.9) & (0.0, 4.8) & (30.3, 32.9) & (10.8, 12.8) \\

& Llama 3.2 11B-V
& (41.6, 47.4) & (26.2, 31.3) & (69.2, 74.6) & (8.9, 13.1) & (4.5, 8.6) & (0.0, 3.9) & (0.0, 2.4) & (30.2, 32.7) & (10.7, 12.7)\\

& Qwen 3 VL 32B
& (45.8, 51.9) & (30.4, 35.9) & (71.0, 76.2) & (10.3, 14.7) & (4.9, 9.3) & (0.0, 6.6) & (0.0, 3.4) & (32.6, 35.2) & (12.4, 14.4) \\

\bottomrule
\end{tabular}
}
\end{table*}
\begin{table*}[t]
\centering
\scriptsize
\setlength{\tabcolsep}{2.3pt}
\renewcommand{\arraystretch}{0.92}
\caption{95\% Confidence Intervals of models on \GD\ validation set, for Highlight Prediction. Models that do not have access to question image are in \textit{{\colorbox[HTML]{E4DFF4}{lavender italics}}}. Confidence Intervals are obtained by bootstrapping the sample with a sample size of $10000$}
\label{tab:results_val_ci2}
\resizebox{\textwidth}{!}{
\begin{tabular}{cl|cccc|cccc}
\toprule
\textbf{Group} &
\textbf{Model} &
\multicolumn{4}{c|}{\textbf{Highlight Prediction Accuracy(\%)}} &
\multicolumn{4}{c|}{\textbf{Highlight Prediction F1(\%)}}  \\
\cmidrule(lr){3-6}
\cmidrule(lr){7-10}

& &
 \textbf{Line} & \textbf{Angle} & \textbf{Arc} & \textbf{Label}& \textbf{Line} & \textbf{Angle} & \textbf{Arc} & \textbf{Label}
\\
\midrule

\multirow{1}{*}{\rotatebox[origin=c]{90}{}}
& Random Turn
& (63.4, 68.6) & (69.2, 74.2) & (91.5, 94.2) & (88.7, 92.0) & (23.0, 32.5) & (12.3, 21.8) & (1.9, 16.3) & (1.5, 12.6) \\

\midrule

\multirow{3}{*}{\rotatebox[origin=c]{90}{\textbf{Closed}}}
& Gemini 2.5 Flash &
(61.4, 66.8) & (67.3, 72.5) & (93.5, 95.9) & (73.6, 78.4) & (45.1, 53.1) & (36.5, 46.1) & (9.9, 33.0) & (9.8, 20.0) \\

& GPT-4o
& (42.4, 47.8) & (68.7, 73.7) & (94.6, 96.9) & (92.3, 95.0) & (41.1, 48.0) & (35.3, 45.2) & (28.6, 54.3) & (0.0, 11.9)\\

& Claude Sonnet 4
& (71.4, 76.2) & (78.0, 82.3) & (96.2, 98.1) & (84.7, 88.4) & (43.8, 53.4) & (31.6, 43.6) & (18.5, 52.0) & (6.5, 19.0) \\

\midrule

\multirow{7}{*}{\rotatebox[origin=c]{90}{\textbf{Open}}}
& {\cellcolor[HTML]{E4DFF4}\textit{Qwen 3 8B}}
& {\cellcolor[HTML]{E4DFF4}\textit{(43.5, 49.0) }}&{\cellcolor[HTML]{E4DFF4}\textit{ (43.2, 48.7) }}&{\cellcolor[HTML]{E4DFF4}\textit{ (92.3, 95.0) }}&{\cellcolor[HTML]{E4DFF4}\textit{ (80.7, 84.9) }}&{\cellcolor[HTML]{E4DFF4}\textit{ (41.9, 48.9) }}&{\cellcolor[HTML]{E4DFF4}\textit{ (32.7, 40.1) }}&{\cellcolor[HTML]{E4DFF4}\textit{ (22.2, 44.2) }}&{\cellcolor[HTML]{E4DFF4}\textit{ (4.9, 15.3)}} \\

& {\cellcolor[HTML]{E4DFF4}\textit{Llama 3.1 8B}}
& {\cellcolor[HTML]{E4DFF4}\textit{(41.1, 46.6) }}&{\cellcolor[HTML]{E4DFF4}\textit{ (74.5, 79.2) }}&{\cellcolor[HTML]{E4DFF4}\textit{ (96.1, 97.9) }}&{\cellcolor[HTML]{E4DFF4}\textit{ (86.7, 90.3) }}&{\cellcolor[HTML]{E4DFF4}\textit{ (39.1, 46.1) }}&{\cellcolor[HTML]{E4DFF4}\textit{ (26.7, 38.2) }}&{\cellcolor[HTML]{E4DFF4}\textit{ (8.7, 40.7) }}&{\cellcolor[HTML]{E4DFF4}\textit{ (0.0, 8.9)}}\\

& {\cellcolor[HTML]{E4DFF4}\textit{Qwen 3 32B}}
& {\cellcolor[HTML]{E4DFF4}\textit{(45.8, 51.3) }}&{\cellcolor[HTML]{E4DFF4}\textit{ (57.4, 62.8) }}&{\cellcolor[HTML]{E4DFF4}\textit{ (92.5, 95.1) }}&{\cellcolor[HTML]{E4DFF4}\textit{ (86.9, 90.5) }}&{\cellcolor[HTML]{E4DFF4}\textit{ (42.1, 49.1) }}&{\cellcolor[HTML]{E4DFF4}\textit{ (37.8, 46.0) }}&{\cellcolor[HTML]{E4DFF4}\textit{ (24.8, 46.8) }}&{\cellcolor[HTML]{E4DFF4}\textit{ (2.9, 15.7)}} \\

& Qwen 3 VL 8B
& (42.3, 47.8) & (63.4, 68.6) & (95.1, 97.3) & (52.5, 58.0) & (40.2, 47.2) & (38.7, 47.5) & (8.7, 37.0) & (8.4, 15.1) \\

& InternVL 8B
& (68.3, 73.3) & (74.1, 78.8) & (95.7, 97.7) & (93.5, 95.9) & (20.8, 31.0) & (11.4, 21.7) & (0.0, 0.0) & (0.0, 0.0) \\

& Llama 3.2 11B-V
& (42.0, 47.6) & (72.0, 76.8) & (95.8, 97.7) & (87.1, 90.6) & (39.3, 46.3) & (23.0, 34.1) & (4.0, 31.1) & (0.0, 9.2) \\

& Qwen 3 VL 32B
& (43.1, 48.6) & (41.9, 47.4) & (95.4, 97.4) & (59.7, 65.2) & (40.6, 47.5) & (33.7, 41.1) & (25.4, 53.3) & (6.2, 13.1)\\

\midrule

\multirow{7}{*}{\rotatebox[origin=c]{90}{\textbf{Fine-tuned}}}
& {\cellcolor[HTML]{E4DFF4}\textit{Qwen 3 8B}}
& {\cellcolor[HTML]{E4DFF4}\textit{(76.0, 80.7) }}&{\cellcolor[HTML]{E4DFF4}\textit{ (79.4, 83.7) }}&{\cellcolor[HTML]{E4DFF4}\textit{ (96.2, 98.1) }}&{\cellcolor[HTML]{E4DFF4}\textit{ (91.1, 93.9) }}&{\cellcolor[HTML]{E4DFF4}\textit{ (42.1, 52.8) }}&{\cellcolor[HTML]{E4DFF4}\textit{ (35.4, 47.6) }}&{\cellcolor[HTML]{E4DFF4}\textit{ (8.9, 42.1) }}&{\cellcolor[HTML]{E4DFF4}\textit{ (9.5, 28.6)}}\\

& {\cellcolor[HTML]{E4DFF4}\textit{Llama 3.1 8B}}
& {\cellcolor[HTML]{E4DFF4}\textit{(76.5, 81.0) }}&{\cellcolor[HTML]{E4DFF4}\textit{ (79.6, 83.9) }}&{\cellcolor[HTML]{E4DFF4}\textit{ (96.2, 98.0) }}&{\cellcolor[HTML]{E4DFF4}\textit{ (91.9, 94.6) }}&{\cellcolor[HTML]{E4DFF4}\textit{ (42.5, 53.3) }}&{\cellcolor[HTML]{E4DFF4}\textit{ (33.2, 45.6) }}&{\cellcolor[HTML]{E4DFF4}\textit{ (5.4, 37.8) }}&{\cellcolor[HTML]{E4DFF4}\textit{ (10.3, 31.2)}}\\

& {\cellcolor[HTML]{E4DFF4}\textit{Qwen 3 32B}}
& {\cellcolor[HTML]{E4DFF4}\textit{(77.2, 81.7) }}&{\cellcolor[HTML]{E4DFF4}\textit{ (80.7, 84.8) }}&{\cellcolor[HTML]{E4DFF4}\textit{ (96.2, 98.0) }}&{\cellcolor[HTML]{E4DFF4}\textit{ (91.9, 94.6) }}&{\cellcolor[HTML]{E4DFF4}\textit{ (42.8, 53.5) }}&{\cellcolor[HTML]{E4DFF4}\textit{ (35.1, 47.7) }}&{\cellcolor[HTML]{E4DFF4}\textit{ (11.1, 44.0) }}&{\cellcolor[HTML]{E4DFF4}\textit{ (5.3, 24.0)}} \\

& Qwen 3 VL 8B
& (77.2, 81.8) & (79.1, 83.5) & (96.1, 98.0) & (91.0, 93.9) & (45.2, 55.6) & (32.2, 44.5) & (13.6, 45.8) & (8.1, 27.0)\\

& InternVL 8B
& (75.6, 80.2) & (80.5, 84.7) & (96.1, 97.9) & (91.0, 93.9) & (43.9, 54.3) & (36.3, 48.9) & (5.4, 37.2) & (9.4, 28.8) \\

& Llama 3.2 11B-V
& (77.6, 82.0) & (79.1, 83.4) & (95.8, 97.8) & (91.1, 94.0) & (43.2, 54.1) & (34.0, 46.2) & (12.8, 44.0) & (3.8, 20.2) \\

& Qwen 3 VL 32B
& (77.8, 82.3) & (80.4, 84.7) & (96.1, 98.0) & (92.6, 95.2) & (45.0, 55.9) & (39.4, 51.7) & (5.3, 37.2) & (7.1, 28.2) \\

\bottomrule
\end{tabular}
}
\end{table*}
\begin{table*}[t]
\centering
\footnotesize
\setlength{\tabcolsep}{2.3pt}
\renewcommand{\arraystretch}{0.92}
\caption{Performance of models on \GD\ validation set. Text-only models are in \textit{{\colorbox[HTML]{E4DFF4}{lavender italics}}}. Best performing models are in bold. See \Cref{tab:results_val_ci1,tab:results_val_ci2} for confidence intervals.}
\label{tab:results_val}
\resizebox{\textwidth}{!}{
\begin{tabular}{clccc|cccc|cccc|cc}
\toprule
\textbf{Group} &
\textbf{Model} &
\multicolumn{3}{c|}{\textbf{Categorical Macro-F1}} &
\multicolumn{4}{c|}{\textbf{Highlight Pred. Acc./F1}} &
\multicolumn{4}{c|}{\textbf{Decision F1}} &
\multicolumn{2}{c}{\textbf{Utterance}} \\
\cmidrule(lr){3-5}
\cmidrule(lr){6-9}
\cmidrule(lr){10-13}
\cmidrule(lr){14-15}
& &
\textbf{Act} & \textbf{Subact} & \textbf{Feedback}
& \textbf{Line} & \textbf{Angle} & \textbf{Arc} & \textbf{Label}
& \textbf{Line} & \textbf{Angle} & \textbf{Arc} & \textbf{Label}
& \textbf{BERT} & \textbf{BLEU} \\
\midrule

\multirow{1}{*}{\rotatebox[origin=c]{90}{}}
& Random Turn
& 27.0 & 15.2 & 46.8 & 66.1/27.8 & 71.7/16.9 & 92.9/8.2 & 90.4/6.2 & 1.0 & 0.3 & 0.0 & 0.8 & 14.1 & 3.7\\

\midrule

\multirow{3}{*}{\rotatebox[origin=c]{90}{\textbf{Closed}}}
& Gemini 2.5 Flash
& 32.1 & 15.4 & 59.1
& 64.1 / 49.2 & 70.0 / 41.4 & 94.7 / 21.4 & 76.0 / 14.8
& 21.5 & 12.9 & 1.3 & \textbf{10.6}
& 15.3 & 3.4 \\

& GPT-4o
& 29.2 & 15.1 & 52.3
& 45.1 / 44.6 & 71.2 / 40.4 & 95.8 / 42.2 & 93.7 / 4.8
& 27.3 & 10.5 & 0.0 & 0.8
& 20.8 & 3.5 \\

& Claude Sonnet 4
& 32.5 & 20.1 & 58.0
& 73.8 / 48.6 & 80.2 / 37.7 & \textbf{97.2} / 36.4 & 86.6 / 12.5
& 15.6 & 7.9 & 0.0 & 3.4
& 25.4 & 6.1 \\

\midrule

\multirow{7}{*}{\rotatebox[origin=c]{90}{\textbf{Open}}}
& {\cellcolor[HTML]{E4DFF4}\textit{Qwen 3 8B}}
& {\cellcolor[HTML]{E4DFF4}\textit{27.5 }}&{\cellcolor[HTML]{E4DFF4}\textit{ 11.5 }}&{\cellcolor[HTML]{E4DFF4}\textit{ 45.9 }}&{\cellcolor[HTML]{E4DFF4}\textit{ 46.2/45.4 }}&{\cellcolor[HTML]{E4DFF4}\textit{ 45.9/36.4 }}&{\cellcolor[HTML]{E4DFF4}\textit{ 93.7/33.6 }}&{\cellcolor[HTML]{E4DFF4}\textit{ 82.8/10.0 }}&{\cellcolor[HTML]{E4DFF4}\textit{ 22.9 }}&{\cellcolor[HTML]{E4DFF4}\textit{ 12.4 }}&{\cellcolor[HTML]{E4DFF4}\textit{ 0.0 }}&{\cellcolor[HTML]{E4DFF4}\textit{ 1.1 }}&{\cellcolor[HTML]{E4DFF4}\textit{ 16.4 }}&{\cellcolor[HTML]{E4DFF4}\textit{ 1.7}} \\

& {\cellcolor[HTML]{E4DFF4}\textit{Llama 3.1 8B}}
& {\cellcolor[HTML]{E4DFF4}\textit{20.5 }}&{\cellcolor[HTML]{E4DFF4}\textit{ 10.3 }}&{\cellcolor[HTML]{E4DFF4}\textit{ 43.8 }}&{\cellcolor[HTML]{E4DFF4}\textit{ 43.8/42.6 }}&{\cellcolor[HTML]{E4DFF4}\textit{ 76.8/32.6 }}&{\cellcolor[HTML]{E4DFF4}\textit{ 97.0/24.5 }}&{\cellcolor[HTML]{E4DFF4}\textit{ 88.6/4.0 }}&{\cellcolor[HTML]{E4DFF4}\textit{ 16.1 }}&{\cellcolor[HTML]{E4DFF4}\textit{ 4.0 }}&{\cellcolor[HTML]{E4DFF4}\textit{ 0.0 }}&{\cellcolor[HTML]{E4DFF4}\textit{ 0.8 }}&{\cellcolor[HTML]{E4DFF4}\textit{ 18.4 }}&{\cellcolor[HTML]{E4DFF4}\textit{ 3.0}} \\

& {\cellcolor[HTML]{E4DFF4}\textit{Qwen 3 32B}}
& {\cellcolor[HTML]{E4DFF4}\textit{30.0 }}&{\cellcolor[HTML]{E4DFF4}\textit{ 12.5 }}&{\cellcolor[HTML]{E4DFF4}\textit{ 50.9 }}&{\cellcolor[HTML]{E4DFF4}\textit{ 48.6/45.6 }}&{\cellcolor[HTML]{E4DFF4}\textit{ 60.1/42.0 }}&{\cellcolor[HTML]{E4DFF4}\textit{ 93.8/36.4 }}&{\cellcolor[HTML]{E4DFF4}\textit{ 88.7/9.0 }}&{\cellcolor[HTML]{E4DFF4}\textit{ 24.5 }}&{\cellcolor[HTML]{E4DFF4}\textit{ 13.9 }}&{\cellcolor[HTML]{E4DFF4}\textit{ 2.6 }}&{\cellcolor[HTML]{E4DFF4}\textit{ 1.9 }}&{\cellcolor[HTML]{E4DFF4}\textit{ 19.3 }}&{\cellcolor[HTML]{E4DFF4}\textit{ 2.9}} \\

& Qwen 3 VL 8B
& 32.2 & 13.3 & 54.3
& 45.0 / 43.8 & 66.1 / 43.3 & 96.2 / 23.0 & 55.2 / 11.7
& 21.0 & 12.6 & 0.0 & 7.2
& 19.8 & 3.2 \\

& InternVL 8B
& 5.0 & 2.5 & 13.0
& 70.8 / 25.9 & 76.5 / 16.5 & 96.7 / 0.0 & \textbf{94.7} / 0.0
& 5.0 & 2.8 & 0.0 & 0.0
& 3.2 & 0.5 \\

& Llama 3.2 11B-V
& 18.6 & 7.2 & 42.0
& 44.8 / 42.8 & 74.4 / 28.6 & 96.8 / 16.7 & 88.9 / 4.1
& 16.5 & 3.4 & 0.0 & 0.8
& 18.3 & 2.8 \\

& Qwen 3 VL 32B
& 30.3 & 14.1 & 56.6
& 45.8 / 44.2 & 44.6 / 37.5 & 96.4 / \textbf{40.0} & 62.5 / 9.6
& 23.9 & \textbf{18.0} & 0.0 & 6.1
& 20.0 & 3.7 \\

\midrule

\multirow{7}{*}{\rotatebox[origin=c]{90}{\textbf{Fine-tuned}}}
& {\cellcolor[HTML]{E4DFF4}\textit{Qwen 3 8B}}
& {\cellcolor[HTML]{E4DFF4}\textit{44.6 }}&{\cellcolor[HTML]{E4DFF4}\textit{ 28.4 }}&{\cellcolor[HTML]{E4DFF4}\textit{ 68.3 }}&{\cellcolor[HTML]{E4DFF4}\textit{ 78.4/47.6 }}&{\cellcolor[HTML]{E4DFF4}\textit{ 81.6/41.6 }}&{\cellcolor[HTML]{E4DFF4}\textit{ 97.2/25.5 }}&{\cellcolor[HTML]{E4DFF4}\textit{ 92.5/19.0 }}&{\cellcolor[HTML]{E4DFF4}\textit{ 12.2 }}&{\cellcolor[HTML]{E4DFF4}\textit{ 7.0 }}&{\cellcolor[HTML]{E4DFF4}\textit{ 0.0 }}&{\cellcolor[HTML]{E4DFF4}\textit{ 4.5 }}&{\cellcolor[HTML]{E4DFF4}\textit{ 31.8 }}&{\cellcolor[HTML]{E4DFF4}\textit{ 12.3}} \\

& {\cellcolor[HTML]{E4DFF4}\textit{Llama 3.1 8B}}
& {\cellcolor[HTML]{E4DFF4}\textit{45.2 }}&{\cellcolor[HTML]{E4DFF4}\textit{ 30.0 }}&{\cellcolor[HTML]{E4DFF4}\textit{ 68.3 }}&{\cellcolor[HTML]{E4DFF4}\textit{ 78.8/48.0 }}&{\cellcolor[HTML]{E4DFF4}\textit{ 81.7/39.6 }}&{\cellcolor[HTML]{E4DFF4}\textit{ 97.1/21.7 }}&{\cellcolor[HTML]{E4DFF4}\textit{ 93.3/20.8 }}&{\cellcolor[HTML]{E4DFF4}\textit{ 12.2 }}&{\cellcolor[HTML]{E4DFF4}\textit{ 5.6 }}&{\cellcolor[HTML]{E4DFF4}\textit{ 0.0 }}&{\cellcolor[HTML]{E4DFF4}\textit{ 1.6 }}&{\cellcolor[HTML]{E4DFF4}\textit{ 30.8 }}&{\cellcolor[HTML]{E4DFF4}\textit{ 11.5}} \\

& {\cellcolor[HTML]{E4DFF4}\textit{Qwen 3 32B}}
& {\cellcolor[HTML]{E4DFF4}\textit{48.1 }}&{\cellcolor[HTML]{E4DFF4}\textit{ 30.6 }}&{\cellcolor[HTML]{E4DFF4}\textit{ 72.3 }}&{\cellcolor[HTML]{E4DFF4}\textit{ 79.5/48.3 }}&{\cellcolor[HTML]{E4DFF4}\textit{ 82.7/41.6 }}&{\cellcolor[HTML]{E4DFF4}\textit{ 97.1/28.0 }}&{\cellcolor[HTML]{E4DFF4}\textit{ 93.3/14.3 }}&{\cellcolor[HTML]{E4DFF4}\textit{ 11.6 }}&{\cellcolor[HTML]{E4DFF4}\textit{ 6.7 }}&{\cellcolor[HTML]{E4DFF4}\textit{ 5.3 }}&{\cellcolor[HTML]{E4DFF4}\textit{ 1.6 }}&{\cellcolor[HTML]{E4DFF4}\textit{ 33.0 }}&{\cellcolor[HTML]{E4DFF4}\textit{ 12.9}} \\

& Qwen 3 VL 8B
& 45.4 & 30.2 & 69.8
& 79.5 / 50.5 & 81.3 / 38.4 & 97.0 / 30.2 & 92.5 / 17.5
& 12.7 & 6.4 & 1.3 & 2.4
& 31.8 & 12.4 \\

& InternVL 8B
& 46.0 & 31.8 & 70.5
& 78.0 / 49.3 & \textbf{82.7} / 42.7 & 97.0 / 21.3 & 92.5 / \textbf{19.0}
& 13.7 & 7.5 & 1.3 & 2.1
& 31.6 & 11.8 \\

& Llama 3.2 11B-V
& 44.5 & 28.7 & 72.0
& 79.9 / 48.8 & 81.2 / 40.2 & 96.9 / 29.1 & 92.6 / 11.4
& 11.0 & 6.5 & 1.3 & 0.8
& 31.5 & 11.7 \\

& Qwen 3 VL 32B
& \textbf{48.8} & \textbf{33.1} & \textbf{73.7}
& \textbf{80.0} / \textbf{50.6} & 82.6 / \textbf{45.8} & 97.0 / 21.3 & 93.9 / 17.4
& 12.4 & 7.0 & 2.6 & 1.3
& \textbf{33.9} & \textbf{13.4} \\

\bottomrule
\end{tabular}
}
\end{table*}
\begin{table*}[t]
\centering
\scriptsize
\setlength{\tabcolsep}{2.3pt}
\renewcommand{\arraystretch}{0.92}
\caption{95\% Confidence Intervals of models on \GD\ test set, for Categorical variables, Decision F1 and Utterance quality. Models that do not have access to question image are in \textit{{\colorbox[HTML]{E4DFF4}{lavender italics}}}. Confidence Intervals are obtained by bootstrapping the sample with a sample size of $10000$}
\label{tab:results_test_ci1}
\resizebox{\textwidth}{!}{
\begin{tabular}{clccc|cccc|cc}
\toprule
\textbf{Group} &
\textbf{Model} &
\multicolumn{3}{c|}{\textbf{Categorical Macro-F1}} &
\multicolumn{4}{c|}{\textbf{Decision F1}} &
\multicolumn{2}{c}{\textbf{Utterance}} \\
\cmidrule(lr){3-5}
\cmidrule(lr){6-9}
\cmidrule(lr){10-11}
& &
\textbf{Act} & \textbf{Subact} & \textbf{Feedback}
& \textbf{Line} & \textbf{Angle} & \textbf{Arc} & \textbf{Label}
& \textbf{BERT} & \textbf{BLEU} \\
\midrule

\multirow{1}{*}{\rotatebox[origin=c]{90}{}}
& Random Turn
&(24.1, 27.0) & (11.9, 13.7) & (45.7, 49.1) & (0.0, 0.4) & (0.0, 0.6) & (0.0, 0.0) & (0.0, 0.6) & (14.7, 16.0) & (3.1, 3.7))\\

\midrule

\multirow{3}{*}{\rotatebox[origin=c]{90}{\textbf{Closed}}}
& Gemini 2.5 Flash
& (28.6, 32.3) & (16.0, 19.0) & (53.5, 57.5) & (16.6, 19.6) & (8.9, 11.8) & (0.4, 3.5) & (2.7, 6.0) & (24.6, 25.8) & (5.0, 5.8) \\

& GPT-4o
& (27.9, 32.2) & (11.9, 15.2) & (49.6, 53.2) & (18.7, 21.6) & (5.6, 8.0) & (0.0, 2.2) & (0.0, 0.2) & (20.9, 22.1) & (3.3, 3.8) \\

& Claude Sonnet 4
& (31.0, 34.4) & (16.4, 19.5) & (51.7, 55.5) & (10.0, 12.8) & (4.7, 7.1) & (0.0, 2.2) & (4.8, 8.8) & (25.1, 26.2) & (5.7, 6.4) \\

\midrule

\multirow{7}{*}{\rotatebox[origin=c]{90}{\textbf{Open}}}
& {\cellcolor[HTML]{E4DFF4}\textit{Qwen 3 8B}}
& {\cellcolor[HTML]{E4DFF4}\textit{(25.3, 28.8) }}&{\cellcolor[HTML]{E4DFF4}\textit{ (9.0, 12.3) }}&{\cellcolor[HTML]{E4DFF4}\textit{ (42.1, 45.5) }}&{\cellcolor[HTML]{E4DFF4}\textit{ (18.6, 21.3) }}&{\cellcolor[HTML]{E4DFF4}\textit{ (9.6, 12.3) }}&{\cellcolor[HTML]{E4DFF4}\textit{ (0.0, 0.0) }}&{\cellcolor[HTML]{E4DFF4}\textit{ (0.8, 2.8) }}&{\cellcolor[HTML]{E4DFF4}\textit{ (16.7, 17.6) }}&{\cellcolor[HTML]{E4DFF4}\textit{ (1.9, 2.3)}} \\

& {\cellcolor[HTML]{E4DFF4}\textit{Llama 3.1 8B}}
& {\cellcolor[HTML]{E4DFF4}\textit{(14.1, 19.7) }}&{\cellcolor[HTML]{E4DFF4}\textit{ (4.0, 6.8) }}&{\cellcolor[HTML]{E4DFF4}\textit{ (43.8, 47.3) }}&{\cellcolor[HTML]{E4DFF4}\textit{ (13.0, 15.7) }}&{\cellcolor[HTML]{E4DFF4}\textit{ (1.5, 3.0) }}&{\cellcolor[HTML]{E4DFF4}\textit{ (0.0, 1.3) }}&{\cellcolor[HTML]{E4DFF4}\textit{ (0.0, 1.0) }}&{\cellcolor[HTML]{E4DFF4}\textit{ (18.6, 19.5) }}&{\cellcolor[HTML]{E4DFF4}\textit{ (2.8, 3.3)}} \\

& {\cellcolor[HTML]{E4DFF4}\textit{Qwen 3 32B}}
& {\cellcolor[HTML]{E4DFF4}\textit{(27.9, 31.2) }}&{\cellcolor[HTML]{E4DFF4}\textit{ (12.1, 16.2) }}&{\cellcolor[HTML]{E4DFF4}\textit{ (49.4, 53.0) }}&{\cellcolor[HTML]{E4DFF4}\textit{ (17.8, 20.6) }}&{\cellcolor[HTML]{E4DFF4}\textit{ (9.6, 12.5) }}&{\cellcolor[HTML]{E4DFF4}\textit{ (0.0, 0.0) }}&{\cellcolor[HTML]{E4DFF4}\textit{ (0.7, 2.9) }}&{\cellcolor[HTML]{E4DFF4}\textit{ (19.8, 20.7) }}&{\cellcolor[HTML]{E4DFF4}\textit{ (2.9, 3.3)}} \\

& Qwen 3 VL 8B
& (28.6, 32.2) & (12.3, 15.0) & (53.0, 56.8) & (15.2, 17.9) & (8.6, 11.4) & (0.0, 3.0) & (3.2, 6.5) & (20.6, 21.6) & (3.3, 3.8) \\

& InternVL 8B
& (3.0, 4.2) & (0.6, 1.4) & (12.1, 15.4) & (2.6, 4.1) & (0.5, 1.5) & (0.0, 0.0) & (0.0, 0.4) & (3.2, 3.8) & (0.5, 0.7) \\

& Llama 3.2 11B-V
& (17.2, 21.4) & (4.5, 7.6) & (42.5, 46.0) & (12.2, 14.7) & (1.6, 3.2) & (0.0, 1.3) & (0.0, 1.4) & (18.5, 19.4) & (2.7, 3.1)\\

& Qwen 3 VL 32B &
(27.7, 31.0) & (11.7, 14.1) & (51.2, 55.6) & (18.8, 21.6) & (11.9, 14.9) & (0.4, 3.5) & (7.8, 12.3) & (20.8, 21.8) & (3.9, 4.4) \\

\midrule

\multirow{7}{*}{\rotatebox[origin=c]{90}{\textbf{Fine-tuned}}}
& {\cellcolor[HTML]{E4DFF4}\textit{Qwen 3 8B}}
& {\cellcolor[HTML]{E4DFF4}\textit{(43.6, 47.2) }}&{\cellcolor[HTML]{E4DFF4}\textit{ (27.0, 30.3) }}&{\cellcolor[HTML]{E4DFF4}\textit{ (69.9, 73.3) }}&{\cellcolor[HTML]{E4DFF4}\textit{ (7.7, 10.2) }}&{\cellcolor[HTML]{E4DFF4}\textit{ (4.5, 6.8) }}&{\cellcolor[HTML]{E4DFF4}\textit{ (0.3, 3.2) }}&{\cellcolor[HTML]{E4DFF4}\textit{ (1.8, 4.7) }}&{\cellcolor[HTML]{E4DFF4}\textit{ (33.3, 34.8) }}&{\cellcolor[HTML]{E4DFF4}\textit{ (12.4, 13.6)}} \\

& {\cellcolor[HTML]{E4DFF4}\textit{Llama 3.1 8B}}
& {\cellcolor[HTML]{E4DFF4}\textit{(42.0, 45.5) }}&{\cellcolor[HTML]{E4DFF4}\textit{ (27.3, 30.3) }}&{\cellcolor[HTML]{E4DFF4}\textit{ (67.3, 70.8) }}&{\cellcolor[HTML]{E4DFF4}\textit{ (7.8, 10.3) }}&{\cellcolor[HTML]{E4DFF4}\textit{ (4.4, 6.6) }}&{\cellcolor[HTML]{E4DFF4}\textit{ (1.7, 6.5) }}&{\cellcolor[HTML]{E4DFF4}\textit{ (1.4, 4.0) }}&{\cellcolor[HTML]{E4DFF4}\textit{ (32.3, 33.9) }}&{\cellcolor[HTML]{E4DFF4}\textit{ (12.1, 13.3)}}\\

& {\cellcolor[HTML]{E4DFF4}\textit{Qwen 3 32B}}
& {\cellcolor[HTML]{E4DFF4}\textit{(43.8, 47.4) }}&{\cellcolor[HTML]{E4DFF4}\textit{ (27.5, 30.4) }}&{\cellcolor[HTML]{E4DFF4}\textit{ (70.1, 73.5) }}&{\cellcolor[HTML]{E4DFF4}\textit{ (8.5, 11.0) }}&{\cellcolor[HTML]{E4DFF4}\textit{ (5.1, 7.6) }}&{\cellcolor[HTML]{E4DFF4}\textit{ (0.6, 4.3) }}&{\cellcolor[HTML]{E4DFF4}\textit{ (0.8, 3.2) }}&{\cellcolor[HTML]{E4DFF4}\textit{ (34.2, 35.7) }}&{\cellcolor[HTML]{E4DFF4}\textit{ (13.4, 14.7)}}\\

& Qwen 3 VL 8B
& (44.5, 48.2) & (29.2, 32.6) & (69.9, 73.4) & (8.2, 10.8) & (7.2, 10.0) & (1.6, 5.7) & (1.6, 4.3) & (33.7, 35.2) & (13.1, 14.5) \\

& InternVL 8B
& (42.2, 45.8) & (27.7, 30.7) & (68.6, 72.1) & (8.5, 11.1) & (5.5, 8.0) & (2.8, 8.1) & (1.3, 3.9) & (33.0, 34.5) & (12.3, 13.5) \\

& Llama 3.2 11B-V
& (41.7, 45.3) & (26.4, 29.1) & (67.4, 70.8) & (7.8, 10.2) & (4.5, 6.8) & (1.5, 6.0) & (1.0, 3.5) & (32.4, 33.9) & (12.1, 13.3)\\

& Qwen 3 VL 32B
& (44.7, 48.4) & (29.4, 33.4) & (70.4, 73.8) & (9.5, 12.3) & (5.9, 8.4) & (4.3, 10.7) & (1.7, 4.7) & (34.6, 36.1) & (13.6, 14.9) \\

\bottomrule
\end{tabular}
}
\end{table*}
\begin{table*}[t]
\centering
\scriptsize
\setlength{\tabcolsep}{2.3pt}
\renewcommand{\arraystretch}{0.92}
\caption{95\% Confidence Intervals of models on \GD\ test set, for Highlight Prediction. Models that do not have access to question image are in \textit{{\colorbox[HTML]{E4DFF4}{lavender italics}}}. Confidence Intervals are obtained by bootstrapping the sample with a sample size of $10000$}
\label{tab:results_test_ci2}
\resizebox{\textwidth}{!}{
\begin{tabular}{cl|cccc|cccc}
\toprule
\textbf{Group} &
\textbf{Model} &
\multicolumn{4}{c|}{\textbf{Highlight Prediction Accuracy(\%)}} &
\multicolumn{4}{c|}{\textbf{Highlight Prediction F1(\%)}}  \\
\cmidrule(lr){3-6}
\cmidrule(lr){7-10}

& & \textbf{Line} & \textbf{Angle} & \textbf{Arc} & \textbf{Label}& \textbf{Line} & \textbf{Angle} & \textbf{Arc} & \textbf{Label}
\\
\midrule

\multirow{1}{*}{\rotatebox[origin=c]{90}{}}
& Random Turn
& (66.5, 69.6) & (71.0, 74.0) & (91.9, 93.7) & (85.1, 87.5) & (23.1, 29.1) & (18.1, 24.3) & (0.0, 5.5) & (4.6, 11.1) \\

\midrule

\multirow{3}{*}{\rotatebox[origin=c]{90}{\textbf{Closed}}}
& Gemini 2.5 Flash
& (66.7, 69.8) & (70.0, 73.1) & (93.0, 94.6) & (85.0, 87.4) & (20.6, 26.5) & (15.7, 21.7) & (1.8, 10.0) & (5.8, 12.7) vc\\

& GPT-4o
& (38.3, 41.6) & (70.7, 73.8) & (93.9, 95.4) & (90.5, 92.4) & (36.2, 40.5) & (33.7, 40.1) & (25.6, 40.3) & (0.7, 6.5) \\

& Claude Sonnet 4
& (70.2, 73.2) & (78.0, 80.8) & (95.2, 96.5) & (82.4, 84.9) & (38.2, 44.2) & (32.4, 39.7) & (34.6, 50.2) & (15.5, 23.5) \\

\midrule

\multirow{7}{*}{\rotatebox[origin=c]{90}{\textbf{Open}}}
& {\cellcolor[HTML]{E4DFF4}\textit{Qwen 3 8B}}
& {\cellcolor[HTML]{E4DFF4}\textit{(40.0, 43.4) }}&{\cellcolor[HTML]{E4DFF4}\textit{ (52.9, 56.3) }}&{\cellcolor[HTML]{E4DFF4}\textit{ (90.2, 92.1) }}&{\cellcolor[HTML]{E4DFF4}\textit{ (76.1, 78.9) }}&{\cellcolor[HTML]{E4DFF4}\textit{ (38.0, 42.3) }}&{\cellcolor[HTML]{E4DFF4}\textit{ (38.3, 43.1) }}&{\cellcolor[HTML]{E4DFF4}\textit{ (30.7, 41.8) }}&{\cellcolor[HTML]{E4DFF4}\textit{ (8.8, 14.9)}} \\

& {\cellcolor[HTML]{E4DFF4}\textit{Llama 3.1 8B}}
& {\cellcolor[HTML]{E4DFF4}\textit{(37.9, 41.2) }}&{\cellcolor[HTML]{E4DFF4}\textit{ (73.7, 76.6) }}&{\cellcolor[HTML]{E4DFF4}\textit{ (94.5, 95.9) }}&{\cellcolor[HTML]{E4DFF4}\textit{ (85.3, 87.6) }}&{\cellcolor[HTML]{E4DFF4}\textit{ (36.1, 40.5) }}&{\cellcolor[HTML]{E4DFF4}\textit{ (24.8, 31.7) }}&{\cellcolor[HTML]{E4DFF4}\textit{ (21.3, 36.6) }}&{\cellcolor[HTML]{E4DFF4}\textit{ (5.8, 12.8)}} \\

& {\cellcolor[HTML]{E4DFF4}\textit{Qwen 3 32B}}
& {\cellcolor[HTML]{E4DFF4}\textit{(44.9, 48.2) }}&{\cellcolor[HTML]{E4DFF4}\textit{ (64.4, 67.6) }}&{\cellcolor[HTML]{E4DFF4}\textit{ (91.2, 93.1) }}&{\cellcolor[HTML]{E4DFF4}\textit{ (85.3, 87.6) }}&{\cellcolor[HTML]{E4DFF4}\textit{ (39.2, 43.6) }}&{\cellcolor[HTML]{E4DFF4}\textit{ (42.7, 48.0) }}&{\cellcolor[HTML]{E4DFF4}\textit{ (35.3, 46.8) }}&{\cellcolor[HTML]{E4DFF4}\textit{ (5.8, 12.8)}} \\

& Qwen 3 VL 8B
& (40.2, 43.6) & (64.9, 68.0) & (94.3, 95.8) & (52.3, 55.7) & (37.0, 41.3) & (40.1, 45.7) & (20.2, 35.6) & (11.6, 16.1) \\

& InternVL 8B
& (69.5, 72.5) & (73.3, 76.2) & (95.5, 96.8) & (91.2, 93.0) & (18.5, 24.5) & (10.6, 16.6) & (0.0, 7.7) & (0.0, 2.4) \\

& Llama 3.2 11B-V
& (39.0, 42.3) & (73.6, 76.6) & (94.6, 96.0) & (86.4, 88.6) & (35.9, 40.1) & (28.7, 35.5) & (23.1, 38.8) & (6.4, 13.8) \\

& Qwen 3 VL 32B
& (40.5, 43.8) & (50.9, 54.2) & (93.3, 94.8) & (58.3, 61.6) & (38.1, 42.4) & (39.1, 43.7) & (34.3, 47.5) & (15.4, 20.3) \\

\midrule

\multirow{7}{*}{\rotatebox[origin=c]{90}{\textbf{Fine-tuned}}}
& {\cellcolor[HTML]{E4DFF4}\textit{Qwen 3 8B}}
& {\cellcolor[HTML]{E4DFF4}\textit{(77.9, 80.6) }}&{\cellcolor[HTML]{E4DFF4}\textit{ (82.4, 84.9) }}&{\cellcolor[HTML]{E4DFF4}\textit{ (95.7, 97.0) }}&{\cellcolor[HTML]{E4DFF4}\textit{ (88.9, 91.0) }}&{\cellcolor[HTML]{E4DFF4}\textit{ (40.5, 47.5) }}&{\cellcolor[HTML]{E4DFF4}\textit{ (44.0, 51.5) }}&{\cellcolor[HTML]{E4DFF4}\textit{ (26.2, 43.7) }}&{\cellcolor[HTML]{E4DFF4}\textit{ (14.2, 24.3)}} \\

& {\cellcolor[HTML]{E4DFF4}\textit{Llama 3.1 8B}}
& {\cellcolor[HTML]{E4DFF4}\textit{(79.3, 82.0) }}&{\cellcolor[HTML]{E4DFF4}\textit{ (82.3, 84.8) }}&{\cellcolor[HTML]{E4DFF4}\textit{ (96.0, 97.2) }}&{\cellcolor[HTML]{E4DFF4}\textit{ (89.7, 91.8) }}&{\cellcolor[HTML]{E4DFF4}\textit{ (41.5, 48.8) }}&{\cellcolor[HTML]{E4DFF4}\textit{ (40.8, 48.6) }}&{\cellcolor[HTML]{E4DFF4}\textit{ (27.7, 45.9) }}&{\cellcolor[HTML]{E4DFF4}\textit{ (13.7, 24.4)}} \\

& {\cellcolor[HTML]{E4DFF4}\textit{Qwen 3 32B}}
& {\cellcolor[HTML]{E4DFF4}\textit{(79.0, 81.6) }}&{\cellcolor[HTML]{E4DFF4}\textit{ (81.4, 84.0) }}&{\cellcolor[HTML]{E4DFF4}\textit{ (96.0, 97.2) }}&{\cellcolor[HTML]{E4DFF4}\textit{ (89.3, 91.2) }}&{\cellcolor[HTML]{E4DFF4}\textit{ (42.5, 49.4) }}&{\cellcolor[HTML]{E4DFF4}\textit{ (40.4, 48.0) }}&{\cellcolor[HTML]{E4DFF4}\textit{ (33.2, 50.9) }}&{\cellcolor[HTML]{E4DFF4}\textit{ (7.4, 16.5)}} \\

& Qwen 3 VL 8B
& (78.4, 81.2) & (82.8, 85.3) & (95.8, 97.1) & (89.2, 91.2) & (42.0, 48.9) & (45.3, 52.8) & (28.1, 46.0) & (11.6, 21.5)\\

& InternVL 8B
& (78.4, 81.1) & (81.2, 83.8) & (95.5, 96.8) & (89.1, 91.1) & (41.9, 48.7) & (41.1, 48.5) & (28.3, 45.3) & (9.8, 19.3) \\

& Llama 3.2 11B-V
& (78.6, 81.3) & (82.1, 84.7) & (96.0, 97.3) & (89.2, 91.2) & (39.3, 46.3) & (41.5, 49.2) & (31.4, 49.3) & (7.5, 16.5) \\

& Qwen 3 VL 32B
& (79.1, 81.8) & (83.1, 85.5) & (96.0, 97.3) & (89.5, 91.5) & (44.5, 51.3) & (43.2, 50.7) & (36.9, 54.0) & (10.6, 20.6) \\

\bottomrule
\end{tabular}
}
\end{table*}

\begin{table*}[t]
\centering
\footnotesize
\setlength{\tabcolsep}{4.0pt}
\renewcommand{\arraystretch}{1.02}
\caption{Ablation study over all combinations of question text, diagram, correct solution, student solution, and fine-tuning.}
\label{tab:abl_full}

\newcolumntype{C}[1]{>{\centering\arraybackslash}p{#1}}

\begin{tabular}{
ccccc
@{\hspace{5pt}}|@{\hspace{5pt}}
cccc
}
\toprule
\multicolumn{5}{c@{\hspace{5pt}}|@{\hspace{5pt}}}{\textbf{Input / Training Setting}} &
\multicolumn{4}{c}{\textbf{Performance (\%)}} \\
\cmidrule(lr){1-5}
\cmidrule(lr){6-9}
\textbf{Question} &
\textbf{Diagram} &
\textbf{Correct} &
\textbf{Student} &
\textbf{SFT} &
\textbf{Act} &
\textbf{Subact} &
\textbf{Feedback} &
\textbf{Utterance} \\
\textbf{Text} &
&
\textbf{Sol.} &
\textbf{Sol.} &
&
\textbf{Macro-F1} &
\textbf{Macro-F1} &
\textbf{Macro-F1} &
\textbf{BLEU} \\
\midrule

\rowcolor[HTML]{EFEFEF}
Yes & Yes & Yes & Yes & Yes & 45.3 & 30.2 & 69.8 & 12.4 \\
Yes & Yes & Yes & Yes & No  & 32.1 & 13.3 & 54.3 & 3.1 \\

\rowcolor[HTML]{EFEFEF}
Yes & Yes & Yes & No  & Yes & 45.0 & 31.3 & 70.4 & 12.5 \\
Yes & Yes & Yes & No  & No  & 32.5 & 12.5 & 59.5 & 3.4 \\

\rowcolor[HTML]{EFEFEF}
Yes & Yes & No  & Yes & Yes & 45.2 & 29.1 & 67.2 & 12.1 \\
Yes & Yes & No  & Yes & No  & 29.9 & 12.2 & 44.7 & 2.9 \\

\rowcolor[HTML]{EFEFEF}
Yes & Yes & No  & No  & Yes & 43.4 & 28.3 & 67.7 & 11.8 \\
Yes & Yes & No  & No  & No  & 33.5 & 12.8 & 49.3 & 3.5 \\

\rowcolor[HTML]{EFEFEF}
Yes & No  & Yes & Yes & Yes & 43.2 & 27.9 & 71.4 & 12.3 \\
Yes & No  & Yes & Yes & No  & 32.5 & 14.4 & 52.2 & 3.2 \\

\rowcolor[HTML]{EFEFEF}
Yes & No  & Yes & No  & Yes & 44.6 & 29.9 & 69.9 & 12.4 \\
Yes & No  & Yes & No  & No  & 33.1 & 14.2 & 56.9 & 3.5 \\

\rowcolor[HTML]{EFEFEF}
Yes & No  & No  & Yes & Yes & 43.7 & 28.5 & 67.3 & 11.8 \\
Yes & No  & No  & Yes & No  & 31.3 & 13.9 & 45.5 & 3.0 \\

\rowcolor[HTML]{EFEFEF}
Yes & No  & No  & No  & Yes & 41.0 & 26.4 & 66.9 & 12.1 \\
Yes & No  & No  & No  & No  & 32.8 & 14.1 & 44.5 & 3.2 \\

\midrule

\rowcolor[HTML]{EFEFEF}
No  & Yes & Yes & Yes & Yes & 42.9 & 28.3 & 70.7 & 12.4 \\
No  & Yes & Yes & Yes & No  & 32.7 & 13.2 & 52.0 & 3.5 \\

\rowcolor[HTML]{EFEFEF}
No  & Yes & Yes & No  & Yes & 44.7 & 31.1 & 72.0 & 12.0 \\
No  & Yes & Yes & No  & No  & 33.7 & 14.0 & 58.5 & 3.5 \\

\rowcolor[HTML]{EFEFEF}
No  & Yes & No  & Yes & Yes & 42.7 & 26.7 & 67.6 & 11.4 \\
No  & Yes & No  & Yes & No  & 31.6 & 13.2 & 45.6 & 3.2 \\

\rowcolor[HTML]{EFEFEF}
No  & Yes & No  & No  & Yes & 44.5 & 29.9 & 66.6 & 11.6 \\
No  & Yes & No  & No  & No  & 29.8 & 12.7 & 53.0 & 3.1 \\

\rowcolor[HTML]{EFEFEF}
No  & No  & Yes & Yes & Yes & 43.5 & 30.1 & 69.4 & 12.1 \\
No  & No  & Yes & Yes & No  & 32.2 & 14.3 & 52.6 & 3.5 \\

\rowcolor[HTML]{EFEFEF}
No  & No  & Yes & No  & Yes & 43.8 & 29.0 & 71.6 & 12.0 \\
No  & No  & Yes & No  & No  & 32.9 & 16.4 & 50.9 & 3.6 \\

\rowcolor[HTML]{EFEFEF}
No  & No  & No  & Yes & Yes & 44.2 & 28.5 & 67.5 & 11.9 \\
No  & No  & No  & Yes & No  & 32.2 & 14.6 & 45.3 & 3.3 \\

\rowcolor[HTML]{EFEFEF}
No  & No  & No  & No  & Yes & 42.6 & 28.8 & 64.6 & 11.6 \\
No  & No  & No  & No  & No  & 31.5 & 14.1 & 51.8 & 3.4 \\

\bottomrule
\end{tabular}
\end{table*}

\begin{table*}[t]
\centering
\scriptsize
\setlength{\tabcolsep}{3.0pt}
\renewcommand{\arraystretch}{1.08}
\caption{Interview questions, probes, rating definitions, sources, and ratings from the five interviewees. ``Source'' maps to the corresponding questions in \citet{martynova-etal-2025-llms}.}
\label{tab:questions}

\newcolumntype{L}[1]{>{\raggedright\arraybackslash}p{#1}}
\newcolumntype{C}[1]{>{\centering\arraybackslash}p{#1}}

\begin{tabular}{
L{3.0cm}
L{3.5cm}
L{3.2cm}
C{0.75cm}
@{\hspace{4pt}}|@{\hspace{4pt}}
ccccc
}
\toprule
\textbf{Question} &
\textbf{Probes} &
\textbf{Rating} &
\textbf{Source} &
\multicolumn{5}{c}{\textbf{Interviewee Rating}} \\
\cmidrule(lr){5-9}
& & & &
\textbf{R1} & \textbf{R2} & \textbf{R3} & \textbf{R4} & \textbf{R5} \\
\midrule

How engaged are your students in math problem discussions?
&
Do students follow what you are saying? Do students participate in the conversation? Did the student initiate reasoning, or mainly respond? Did the student elaborate explanations? Did the student ask clarifying questions?
&
\textbf{Student Engagement} \newline
1 = AI student engages more; 5 = human students engage more
&
1+2
& 3 & 2 & 2 & 2 & 2 \\

\midrule

What emotions are common to your students due to math confusion?
&
How closely was it represented in the Geo2Dial study? How do you behave when students convey the emotions you listed? Did the student's tone match the mathematical difficulty?
&
\textbf{Student Emotions} \newline
1 = AI student shows more emotions; 5 = human students show more emotions
&
6
& 5 & 5 & 3 & 4 & 3 \\

\midrule

How well did the students understand your diagrams?
&
If instead you were drawing on a board, would things be different? How often do students propose changes to diagrams?
&
\textbf{Diagram Interaction} \newline
1 = AI student understands better; 5 = human students understand better
&
Custom
& 3 & 4 & 1 & 4 & 3 \\

\midrule

What was the common reason for confusion in Geo2Dial?
&
Errors were generated by AI. Do you think it did a good job?
&
\textbf{Error Realism} \newline
1 = errors are very realistic; 5 = errors are not realistic
&
7
& 1 & 1 & 1 & 2 & 2 \\

\midrule

Did you feel frustrated with any of the student conversations?
&
Do students sometimes feel too dense?
&
\textbf{Teacher Frustration} \newline
1 = humans frustrate me more; 5 = AI frustrates me more
&
3
& 2 & 5 & 4 & 3 & 1 \\

\midrule

Did you adjust your teaching strategies in Geo2Dial?
&
Was this in response to our instructions or student behavior? Did you give more or less feedback?
&
\textbf{Strategy Difference} \newline
1 = very little change; 5 = a lot of change
&
4+5+8+9
& 2 & 2 & 1 & 3 & 2 \\

\bottomrule
\end{tabular}
\end{table*}

\section{Case Study}
\label{sec:qual-case-study}

To better understand how supervised fine-tuning changes model behavior, two researchers manually inspected 100 randomly sampled examples from the test set and compared the next-teacher utterances generated by the original and fine-tuned models. We focused on this field because it directly reflects the model's ability to generate pedagogically appropriate tutoring responses. The analysis revealed a consistent behavioral shift: the fine-tuned model was less likely to produce generic geometry explanations and more likely to generate teacher-like tutoring moves that were aligned with the dialog context.

\paragraph{From explanation-heavy responses to pedagogical scaffolding.}

A prominent pattern is that the original model often produces long, explanation-heavy responses. Even when the gold teacher utterance only provides brief feedback, asks the student to continue reasoning, or closes the conversation, the original model tends to restate the relevant geometry rule or explicitly reveal the answer. While these responses are often informative, they can be pedagogically misaligned because they reduce opportunities for student reasoning. In contrast, the fine-tuned model more often produces concise scaffolding moves that better match the local tutoring context, such as acknowledging a correct step, gently correcting an incorrect one, or asking the student to inspect a specific part of the diagram.

\paragraph{Improved dialog-flow management.}

Fine-tuning also improves the model's ability to manage the flow of the tutoring interaction. The original model frequently treats different tutoring stages in a similar way, continuing to explain the geometry regardless of whether the target teacher move is an introduction, a probing question, a hint, or a farewell. For example, in several closing turns, the original model continues to justify the final answer with additional geometric reasoning, whereas the fine-tuned model produces teacher-like closing utterances that praise the student's work and invite future questions. Similarly, for introductory turns, the fine-tuned model is more likely to ask the student to explain their approach before giving targeted guidance, while the original model often jumps directly to a specific mathematical assumption or error. This suggests that fine-tuning helps the model learn not only what mathematical content to discuss, but also when and how to deliver it within a tutoring dialog.

\paragraph{Stronger visual grounding.}

We further observe that the fine-tuned model shows stronger visual grounding. Its utterances more frequently refer to highlighted or visually salient parts of the diagram, using phrases such as ``the angle marked in green,'' ``the highlighted part,'' or ``the figure.'' This behavior is important in our setting because effective geometry tutoring often depends on directing the student's attention to specific diagram components. By contrast, the original model more often discusses geometry concepts abstractly, without explicitly connecting its feedback or question to the visual evidence in the diagram. This difference suggests that fine-tuning encourages the model to imitate visually grounded teacher behavior, rather than only producing generic geometry explanations.

\paragraph{Remaining limitations.}

The case study also reveals several remaining limitations. The fine-tuned model sometimes becomes formulaic, repeatedly using common teacher-like phrases such as ``That's right,'' ``Not exactly,'' or ``Can you walk me through...''. In addition, while its responses are often better aligned with the dialog context, they can occasionally be less conceptually specific than those of the original model. For instance, the fine-tuned model may ask the student to compare two highlighted regions, while the original model more explicitly names the relevant theorem. This suggests that fine-tuning improves conversational and pedagogical alignment, but future work is needed to preserve conceptual specificity and reduce template-like generation.

\section{Interface}
\begin{figure}
    \centering
    \includegraphics[width=\textwidth]{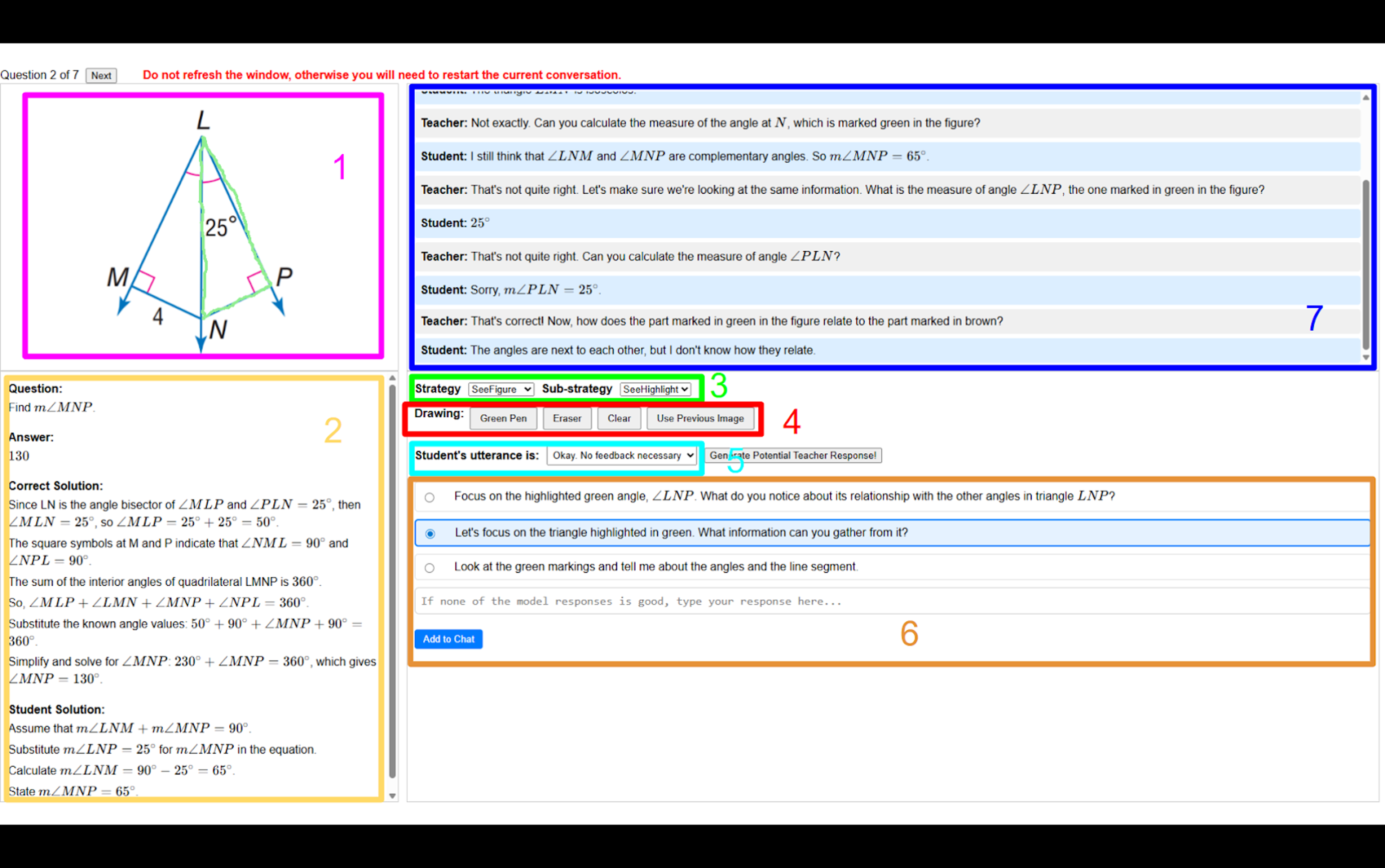}
    \caption{Our annotation interface. Highlighted elements indicate 1. The diagram which can be drawn upon, 2. Question, Answer, Correct Solution and Student Solution, 3. Act and subact selection, 4. Drawing controls, 5. Feedback selection, 6. Teacher utterance suggestions and 7. conversation so far}
    \label{fig:interface}
\end{figure}

\begin{figure}
    \centering
    \includegraphics[width=0.7\linewidth]{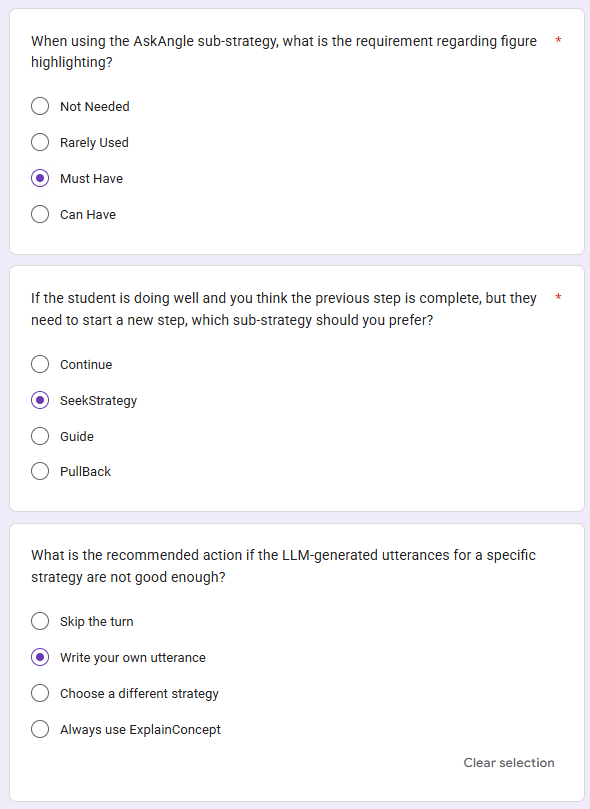}
    \caption{Quiz questions for filtering out annotators who did not do the onboarding properly. Selected options are the correct ones. The associated video can be fount here:\href{https://youtu.be/xtSJ8tPUBCw}{https://youtu.be/xtSJ8tPUBCw}}
    \label{fig:quiz}
\end{figure}

\Cref{fig:interface} shows our annotation interface. \Cref{fig:quiz} shows the annotator onboarding quiz.

\section{Training Prompt}
\label{app:tp}

We used the following prompts for training in  \Cref{sec:7}. 

\begin{lstlisting}
You are given a geometry problem, its associated image, its correct solution, a student's incorrect solution, and a partial tutoring conversation between a tutor and a student where the tutor helps the student to solve the problem. The goal of the tutor is to guide the student towards the correct solution by asking leading questions and providing hints, without directly giving away the answer. Your task is give the next tutor utterance.
Before outputting your response, you need to output the Act & Subact that the teacher should take next, what highlights the teacher should make to the image (if any), and the kind of feedback the teacher should give. We shall now describe the taxonomies for each of these components:

Acts and subacts:
 There are a total of 5 acts and 16 subacts. The acts and their corresponding subacts are as follows:
Act: Generic - Non pedagogical acts not covered by any other category
    Subact: Introduce - Start a new conversation, ask student to explain their approach
    Subact: Continue - Filler utterance, get the student to continue their explanation
    Subact: Farewell - End the conversation politely
Act: Focus - Utterances that take a direct approach towards solving the problem
    Subact: SeekStrategy - Passively pushing the student forward in their problem-solving process
    Subact: Guide - Actively pushing the student forward in their problem-solving process
    Subact: Calculate - Ask the student to make calculations that they proposed in their previous utterance, potentially with a calculator
    Subact: PullBack - Ask the student to start fresh
Act: SeeFigure - Utterances that ask the student to look at the figure, or parts of it. All these utterances need highlights in the figure
    Subact: SeeHighlight - Ask the student to look at the figure. Use this if None of the other options apply. Use green highlight
    Subact: GetRelation - Ask the student to describe the relation between two highlighted parts. One highlight is green, the other is brown
    Subact: AskLength - Ask the student to find the length of the marked line segment or arc. Use green highlight
    Subact: AskAngle - Ask the student to find the angle between two highlighted lines. Use green highlight
Act: Probing - Utterances that ask the student to explain their reasoning or approach, or take more creative but indirect approaches to solving the problem
    Subact: AskForExplanation - Ask the student to explain their reasoning behind a specific step or decision
    Subact: SelfCorrect - Encourage the student to self-correct their approach or reasoning
    Subact: SeekKnowledge - Make the student recall information relevant to the question that they should already know, eg theorem or formulae
Act: Telling - Utterances that reveal part of the solution or provide direct information to the student
    Subact: Reveal - Reveal part of the solution to the student to help them make progress
    Subact: ExplainConcept - Explain a concept or strategy that is relevant to the problem at hand
Give your output as Act: <act_name>, Subact: <subact_name>



Highlighting Instructions:
When highlighting parts of the image, you can choose to highlight lines, angles, arcs, labels.
- Lines should be specified by the two endpoints, e.g., "AB".
- Angles should be specified by the three points, e.g., "ABC" for angle at B formed by points A and C.
- Arcs should list all points appearing one the arc followed by the center point, e.g., "ABCO" for arc from A to C passing through B, and center O.
- Labels can be any alphanumeric character present in the image, or something not covered by the above categories, eg parallel markers on lines, equal markers on angles etc.
Highlight can be mage in green or brown colors. Use brown color only if using green is not sufficient to convey the information.
When specifying highlights, use the following format:
Highlights:
line_green: <line1>;<line2>
angle_green: <angle1>
arc_green: <arc1>
label_green: <label1>
line_brown: <line3>
angle_brown:
arc_brown: <arc2>
label_brown:
ie use semicolons to separate multiple highlights of the same type, and use new lines to separate different types of highlights. If any category has no highlights , leave it blank after the colon.


Feedback Instructions:
Feedback should be positive, negative, or neutral.
- Positive feedback acknowledges correct steps or understanding by the student. use it generously when the student is making progress.
- Negative feedback points out mistakes or misconceptions in the student's approach. Use it only if the student is going off track
- Neutral feedback is used when the student's response is neither particularly good nor bad, or when the tutor is transitioning between topics.
When specifying feedback, use the following format:
Feedback: <Positive/Negative/Neutral>


The overall output format is as follows:
Act: <act_name>, Subact: <subact_name>
Highlights:
line_green: <line1>;<line2>
angle_green: <angle1>
arc_green: <arc1>
label_green: <label1>
line_brown: <line3>
angle_brown:
arc_brown: <arc2>
label_brown:
Feedback: <Positive/Negative/Neutral>
Teacher's next utterance: <utterance>
Make sure to follow the format exactly. Do not output anything else.
\end{lstlisting}

Example user message(image, for models that use it, is supplied separately.)

\begin{lstlisting}
    Question: In \odot D, CF = 8, DE = FD, and DC = 10. Find the measure of AB.
Diagram:<image>Correct solution:
Since DF is perpendicular to chord BC, F is the midpoint of BC, so BC = 2 * CF = 2 * 8 = 16.
In right triangle DFC, use the Pythagorean theorem: DF*DF + CF*CF = DC*DC. Substitute DC = 10 and CF = 8 to find DF = 6.
Given DE = FD, we have DE = 6.
Since chords AB and BC are equidistant from the center D (DE = DF = 6), their lengths are equal.
Therefore, AB = BC = 16.
Student's incorrect solution: Identify D as the center of the circle.
Recognize that DE and DF are radii, therefore DE = DF.
Conclude that a direct calculation of AB is challenging given the provided information.
Select 6.0 as the most reasonable and likely answer.
Conversation so far:
Teacher: Can you walk me through how you approached this problem? I'd like to understand your thinking process from the beginning.
Student: Okay. I identified D as the center of the circle. Then I saw that DE and DF are radii, therefore DE = DF. After that, I realized calculating AB directly would be hard. So I guessed 6.0.
Teacher: That's not quite right. Guessing isn't usually the best strategy in geometry, as we can often find the exact answer with the right tools.

You correctly identified D as the center. Now, let's look at DE and DF. What does a segment from the center that is perpendicular to a chord do to that chord?
Student: It bisects the chord.
Teacher: That's right! Now, let's look at the information that DE = FD. What does it mean for two chords if they are equidistant from the center of the circle?
Student: They are congruent.
Teacher: That's right! Chords equidistant from the center are indeed congruent. Now, let's put that together with the other information we have. What does that tell you about chords AB and BC? Can you use that to find the length of AB?
Student: Since DE = DF, then AB = BC. But I still don't see how to find the length of AB.
Teacher: Good job! You're absolutely right that AB = BC. Now, let's focus on chord BC. We know CF = 8. Do you remember the theorem about a radius or diameter perpendicular to a chord?
Student: Yes, it bisects the chord.
Teacher: Exactly! Now, since DF is perpendicular to chord BC and bisects it, what does that tell you about the relationship between CF and the full length of chord BC? Can you recall the theorem about the perpendicular from the center to a chord?
Student: Since DF bisects BC, then BC = 2 * CF. So, BC = 2 * 8 = 16.
Teacher: Excellent! You've got it - BC = 16. Now, since you already established that AB = BC (because they're equidistant from the center), what does that tell you about the length of AB?
Student: Then AB = 16.

\end{lstlisting}
\section{Student and Teacher Prompts}
\label{app:n}
The following is the syste prompt for the student model:
\begin{lstlisting}
You are a high school student. You tried to solve a geometry problem, but got the wrong answer. You are now talking to a teacher who is trying to help you understand your mistake and learn from it. The problem and your attempt to solve it is as follows:
Problem:{problem}
Your solution:{solution}

Interact with the teacher and work on the problem. Keep your utterances short. Do not budge from your original solution unless the teacher clearly gives you a reason to do so.
\end{lstlisting}
The utterances hereafter are fed in chat format, with teacher utterances used as user messages.

The teacher's suggestions are generated using a combination of a system prompt and a user prompt. The system prompt is as follows:
\begin{lstlisting}
You are a teacher helping a high school student understand their mistake in solving a geometry problem. The student has provided their solution, which is incorrect. Your goal is to guide the student towards understanding their mistake and learning from it. The problem and the student's solution are as follows:
Problem:{problem}
Student's solution:{solution}

The correct answer is {answer}

So far the conversation has gone as follows:
{conversation}

Give the next teacher utterance in the conversation, faithful to the instruction given below. Give only the utterance, no need to append \"Teacher: \" to the beginning of the utterance. Keep the utterance short
\end{lstlisting}
 To this we add the following user prompt:
 \begin{lstlisting}
The figure for the question is attached. Note that the user may have made some annotations on it with green or brown ink. If so, assume that you(the teacher) have made these annotations for the student. Also these annotations may have changed from the previous utterance.
Your goal for the next utterance is:{goal}
{feedback}
 \end{lstlisting}
Here `goal' is replaced by a description of the selected subact. `feedback' is dependent on selected feedback type. It is left empty if feedback is neutral. If positive, it becomes:
\begin{lstlisting}
The students previous utterance was correct. Before proceeding to your goal, give positive feedback to the student, eg "Good job!" or "That's right!"
\end{lstlisting}
and for negative it becomes:
\begin{lstlisting}
The students previous utterance was incorrect. Before proceeding to your goal, give negative but polite feedback to the student, eg "That's not quite right" or "Not Exactly"
\end{lstlisting}
\end{document}